\documentclass[11pt,a4paper]{article}
\usepackage{jheppub}
\usepackage{upquote} 

\usepackage{amsmath,amssymb,amsfonts,dsfont,mathtools,comment} 

\usepackage{placeins}
\usepackage{tabularx,booktabs}
\usepackage[small]{caption}
\usepackage{lscape}
\newcolumntype{M}[1]{>{\centering\arraybackslash}m{#1}}

\usepackage{graphicx}
\graphicspath{{./figures/}}

\newcommand{\ri}{{\rm i}}

\newcommand{\dd}{\mathop{}\!\mathrm{d}}
\allowdisplaybreaks 

\title{NLO in the large charge sector of the critical $O(N)$ model at large $N$ }

\author[a]{Nicola Andrea Dondi}
\author[b]{Giacomo Sberveglieri}

\affiliation[a]{ICTP, Strada Costiera 11,  I-34151 Trieste, Italy\\ INFN, Sede Centrale Via Enrico Fermi 54, Frascati, Italy}
\affiliation[b]{Albert Einstein Center for Fundamental Physics,
Institute for Theoretical Physics, University of Bern,
Sidlerstrasse 5, CH-3012 Bern, Switzerland}

\emailAdd{ndondi@ictp.it}
\emailAdd{giacomo.sberveglieri@unibe.ch}

\abstract{
We compute the next-to-leading correction to the scaling dimension of large-charge operators in the $3d$ critical $O(N)$ model in a double scaling limit in which both $N$ and the operator charge $Q$ are taken to be large. When $Q \gg N$ our result matches predictions from the conformal superfluid EFT and allows to extract next-to-leading order corrections to the EFT Wilsonian coefficients. At present, our result represents the most precise determination of large-charge operator scaling dimension in weakly-coupled CFTs. 
}

\begin{document}

\maketitle

\section{Introduction}
\label{sec:intro}

Sectors of conformal field theories in $d$ dimension corresponding to heavy operators underwent extensive studies in recent years. The underlying motivation is that one has access to non-trivial information concerning these sectors which do not rely on the CFT to be weakly coupled. In unitary CFTs, notable examples are large-spin operators \cite{Fitzpatrick:2012yx, Komargodski:2012ek, Alday:2015eya}, large-global-symmetry quantum numbers operators \cite{Hellerman:2015nra, Monin:2016jmo} but also generic heavy operators away from unitarity bounds \cite{Pappadopulo:2012jk, Mukhametzhanov:2018zja}, with more recent progress based on thermal EFT constructions \cite{Benjamin:2023qsc, Benjamin:2024kdg}. 

In this work, we will focus on heavy operators with large quantum numbers relative to some Cartan generator of the global symmetry group $G$ of the CFT. For the prototypical case $G= U(1)$, it is known that correlators of the form
\begin{equation}
\langle \bar{\mathcal{O}}_{Q \gg 1}(\infty) \dots\mathcal{O}_{Q\gg 1}(-\infty) \rangle_{\mathbb{R} \times S^{d-1}} \,,
\label{eq:large_charge_obs}
\end{equation}
where $\dots$ stems for light small-charge insertions, can be computed on the cylinder using the EFT for a conformal superfluid \cite{Hellerman:2015nra}. Such EFT describes the low-energy dynamics of the Goldstone boson (GB) related to the $U(1)$-breaking on the cylinder.\footnote{More precisely, it is the GB corresponding to the breaking of a combination of the $U(1)$ generator and the cylinder time translation.}\footnote{Strictly speaking, there is no spontaneous symmetry breaking in finite volume, but it is possible to show that the overlap between states in different superselection sectors is heavily suppressed $\sim e^{-Q}$ in the $Q\rightarrow \infty$ limit.} The EFT is built systematically as a derivative expansion in the combination $\chi = -i  \mu \tau + \pi$ comprising of a background configuration\footnote{We denote as $\tau$ the time coordinate on the cylinder, $\mu$ is a dimensionful constant.} and the GB fluctuation $\pi$. The first few terms are
\begin{equation}
	S_{\rm EFT} = \int \dd^d x \sqrt{\hat{g}} \left(  - c_1  + c_2 \hat{\mathcal{R}} - c_3 \hat{\mathcal{R}}_{\mu\nu} \partial^\mu \chi \partial^\nu \chi + c_4 \hat{\mathcal{R}}^2 + \dots\right)\,,
	\label{eq:EFT}
\end{equation}
where all the curvature invariants are computed using the combination $\hat{g}_{\mu\nu} = g_{\mu\nu} (-\partial_\rho \chi \partial^\rho \chi)$ with $g_{\mu\nu}$ the standard cylinder metric. The scale at which the EFT break is set by $\mu$, which can be identified with a chemical potential sourcing the $U(1)$ charge. The Wilson parameters $c_1, c_2, c_3, c_4 ...$ are generically $O(1)$ constants specific to the CFT under consideration and are inaccessible within the EFT. The EFT can be used to compute correlators of the type \eqref{eq:large_charge_obs}, in the special case in which no light insertions are present, one can compute the scaling dimension $\Delta(Q)$ of the $\mathcal{O}_Q$ operator in a power series expansion in $Q$ which controls perturbation theory in the EFT. The result obtained from the ground state configuration $\chi |_{\text{GS}} = - i \mu \tau$ only leads to the prediction
\begin{equation} \label{eq:GS_contribution}
\Delta(Q)|_{\text{GS}} = Q^{\frac{d}{d-1}} \left[ \alpha_{\frac{d}{d-1}} + \alpha_\frac{d-2}{d-1} Q^{- \frac{2}{d-1}} + \alpha_{\frac{d-4}{d-1}} Q^{-\frac{4}{d-1}} + \dots\right],
\end{equation} 
while the (one-loop) quantum corrections introduce new terms of the type \cite{Cuomo:2020rgt}
\begin{equation}
\Delta(Q)= \Delta(Q)|_{\text{GS}} + \left\{ \begin{matrix}
& \beta_0 + \beta_{-\frac{2}{d-1}} Q^{-\frac{2}{d-1}}  +  \dots & d = \text{odd} \\ & \beta_0 \log Q  +\beta_{-\frac{2}{d-1}} Q^{-\frac{2}{d-1}}  \log Q+  \dots\,\,  &  d= \text{even} 
\end{matrix} \right\}\,.
\label{eq:EFT_prediction}
\end{equation}
This expansion will gain a new set of powers (and logs) coming from higher loop corrections \cite{Dondi:2022wli}. The coefficient $\beta_0$ is directly related to the Casimir energy of the conformal GB, in $d=3$ it was found to be \cite{Hellerman:2015nra, Monin:2016bwf}:
\begin{equation} \label{eq:beta0_Casimir}
\beta_0 = \left. \frac{1}{2\sqrt{d-1}} \sum_{\ell=1}^\infty \text{deg}(\ell) \sqrt{\ell(\ell+d-2)} \right|_{\zeta-\text{reg}} \xrightarrow[d \rightarrow 3]{} - 0.093725\dots \, ,
\end{equation}
which is universal, i.e. does not depend on any Wilson coefficient $c_i$ of the EFT. Higher coefficients, like $\beta_{-1}$, are not universal and depend on the Wilson coefficients entering in the modification of the GB dispersion relation \cite{Cuomo:2020rgt}. Some of these results we will use in this work are reviewed in more detail in appendix \ref{sec:EFT}. 
Novelties and modifications of the EFT appear if one considers operators in large asymmetric representations of more complicated global groups $G$ \cite{Hellerman:2017efx, Hellerman:2018sjf} or operators carrying large spin \cite{Cuomo:2019ejv, Cuomo:2022kio}, see also ref.~\cite{Gaume:2020bmp} for a review. For the sake of our work, the EFT (\ref{eq:EFT}) is sufficient. 

The EFT approach for the computation of correlators \eqref{eq:large_charge_obs} is a bottom-up approach: it does not depend on the details of the CFT considered but relies on assumptions about the type of large-charge phase realized on the cylinder. It is generically expected that bosonic CFTs will have large-charge sectors described by the superfluid EFT, but fermionic CFTs might develop phases more similar to Fermi liquids \cite{Komargodski:2021zzy, Dondi:2022zna}. A more top-down approach is required in order to validate such assumptions or to discover entirely new large-charge phases. This can be carried out by sticking to specific models which are weakly coupled with some small controlling parameters. In these cases, the deviation from standard perturbation theory is revealed by taking the appropriate double-scaling limit in which a combination of the charge $Q$ and the small perturbative parameter is kept fixed. At present, there are numerous results for models in the $\epsilon$-expansion \cite{Badel:2019oxl,Antipin:2020abu,Antipin:2020rdw,Antipin:2021jiw,Antipin:2022naw}, in large-$N$ expansions \cite{Alvarez-Gaume:2019biu, Giombi:2020enj, Dondi:2021buw, Dondi:2022zna}, but also numerical Monte Carlo results \cite{Cuomo:2023mxg,Banerjee:2017fcx,Banerjee:2019jpw, Banerjee:2021bbw} and Bootstrap \cite{Rong:2023owx}. In this work, we consider large-charge correlators in the $O(N)$ critical model in dimension $2<d<4$ first analyzed in ref.~\cite{Alvarez-Gaume:2019biu} using a double scaling limit
\begin{equation}
	Q,N \rightarrow \infty, \qquad \hat{Q}\coloneqq Q/N \,\,\text{fixed}. 
\end{equation}
In this model, the lightest large-charge operator can be identified with $\mathcal{O}_Q = \phi_{\{ i_1 } ... \phi_{i_Q\}}$ where $\phi$ is the $O(N)$-fundamental field. Its scaling dimension is expanded in this limit as
\begin{equation}
	\Delta(Q)= N \Delta_{-1}( \hat{Q} ) + N^0 \Delta_0 ( \hat{Q} ) + \dots\,,
	\label{eq:LargeN_Delta}
\end{equation}
where $\Delta_{-1}, \Delta_{0}$ are full functions of $\hat{Q}$ which reproduce the EFT expansion \eqref{eq:EFT_prediction} in the $\hat{Q} \rightarrow \infty$ limit, and interpolate to ordinary large-$N$ perturbation theory in the $\hat{Q} \rightarrow 0$ limit. The main result of this paper is the (numerical) computation of the function $\Delta_0$ in $d=3$, for which we will use an algorithm adapted from a series of works computing large-$N$ scaling dimension of monopole operators \cite{Dyer:2015zha, Pufu:2013eda, Pufu:2013vpa, Chester:2021drl, Chester:2017vdh}. This will allow us to obtain next-to-leading-order (NLO) estimates for the EFT coefficients $c_1, c_2,$ and $c_4$, which we will compare with existing literature, as well as provide a non-trivial check for the value of $\beta_0$ and a relation between $\beta_{-1}$ and the coefficients $c_2,c_3$. A similar methodology was used to compute the coefficient $\beta_0$ in scalar $\text{QED}_3$ in a large-charge monopole background and shown to match the EFT prediction for a conformal GB \cite{DeLaFuente:2018uee}. As a side result, we will also obtain numerical estimates for (parts of) the 3 and 4-loop contribution to the scaling dimension $\Delta(Q)$ in large-$N$ perturbation theory corresponding to the next-to-maximal power in $Q$. 

The plan of this paper is as follows. In section \ref{sec:setup_LO}, we review the leading-order (LO) computation for large-charge operators in the critical $O(N)$ model. In section \ref{sec:NLO}, we present the NLO computation, comment on the $c_3$ coefficient of the EFT and its contribution to the GB dispersion relation, and provide our estimates for the corrections to the coefficients $c_1, c_2,$ and $c_4$. Finally, in section \ref{sec:conclusion}, we present our conclusion and possible future directions. 
We include a series of appendices containing technical details which were not included in the main text.

\section{Setup and leading order computation}
\label{sec:setup_LO}

We follow the approach set up in ref.~\cite{Alvarez-Gaume:2019biu} and compute the ground state energy of the critical $O(N)$ model at finite chemical potential for a Cartan generator on $\mathbb{R} \times S^{d-1}$. This was shown to be analog to the computation of the scaling dimension $\Delta(Q)$ of the operator $\mathcal{O}_Q = \boldsymbol{\phi}_{\{i_1} ... \boldsymbol{\phi}_{\, i_Q\}}$ \cite{Giombi:2020enj}. Here we denote $\boldsymbol{\phi} = ( \phi_1 , ... , \phi_N)$ as the fundamental $O(N)$-field. The bare $O(N)$-model action in dimension $2< d < 4$ in the presence of a chemical potential reads
\begin{align}
	S =& \int \dd^d x \left\{  \frac{1}{2} (\partial \boldsymbol{\phi})^2 + \frac{1}{2} m^2 \boldsymbol{\phi}^2 + \frac{1}{2\sqrt{N}} \sigma \boldsymbol{\phi}^2 - \frac{\sigma^2}{4g} \right\} \nonumber\\
	&+ \int \dd^d x \left\{  \ri \mu (\phi_1 \partial_0 \phi_2 - \phi_2 \partial_0 \phi_1) - \frac{\mu^2}{2} \left[ (\phi_1)^2 + (\phi_2)^2 \right]  \right\} \, ,
	\label{eq:O(N)_action}
\end{align}
where the chemical potential $\mu$ has been chosen, without loss of generality, to be aligned to the generator of $\mathfrak{so}(N)$ which rotates $\phi_1, \phi_2$ only. 
In the above action, parameters and fields have to be consider as bare quantities, renormalized as follows
\begin{equation}
	\boldsymbol{\phi} \rightarrow Z_\phi^{\frac{1}{2}} \boldsymbol{\phi}, \quad m^2 \rightarrow \frac{Z_{m^2}}{Z_\phi} m^2 , \quad \sigma \rightarrow \frac{Z_\sigma}{Z_\phi} \sigma, \quad g \rightarrow \frac{Z_\sigma^2}{Z_g Z_\phi^2} g\,. 
\end{equation}
The perturbative renormalization is carried in a $1/N$ expansion, and in the normalization chosen the coupling $g$ is kept fixed. Renormalization constants are generically in the form $Z_\alpha = 1 + \sum_k \delta Z_{\alpha,k} / N^k$. When computing extensive quantities as free energy or effective potential, it is more convenient to work in terms of bare quantities with some regulator and renormalize the result at the very end. Generically, one expects that at large $N$ path integrals computed with the action \eqref{eq:O(N)_action} will be dominated by homogeneous field saddle point configurations.\footnote{The homogeneity assumption can be relaxed, but in our context it is well justified by the fact that the operator $\mathcal{O}_Q$ is a scalar.} We then expand fields as follows:
\begin{align}
	\sigma &= \sqrt{N} \overline{\Sigma} +  \hat{\sigma} , & &  \boldsymbol{\phi} = \sqrt{N} \left( \overline{\Phi}, 0, ... 0 \right) +  \left( \hat{\phi}_1 , \hat{\phi_2} , \hat{\eta}_1 , ... ,\hat{\eta}_{N-2}\right), \nonumber\\
	\overline{\Sigma}  &= \Sigma+ \sum_{n=1}^\infty \Sigma^{(n)} N^{-n} , & & \overline{\Phi} = \Phi+ \sum_{n=1}^\infty \Phi^{(n)} N^{-n}\,.
\end{align}
The Lagrangian density corresponding to the action above then assumes the appropriate form to develop the $1/N$ expansion:\footnote{We do not expand the VEVs of the field for the moment. }
\begin{equation}
	\mathcal{L}_R = N \mathcal{L}_{\rm tree} + \sqrt{N} \mathcal{L}_{\rm tad} + \mathcal{L}_{\rm gauss} + \frac{1}{\sqrt{N}} \mathcal{L}_{\rm int}\,,
\end{equation}
where one defines 
\begin{align}
	\mathcal{L}_{\rm tree} =& \frac{1}{2} \overline{\Phi}^2 \left(  m^2 - \mu^2 + \overline{\Sigma} \right) -  \frac{\overline{\Sigma}^2}{4g}\,, \nonumber\\
	\mathcal{L}_{\rm tad} =&  \left(m^2 - \mu^2 + \overline{\Sigma} \right) \overline{\Phi} \hat{\phi}_1 + \frac{1}{2} \left( \overline{\Phi}^2-\frac{\overline{\Sigma}}{g} \right) \hat{\sigma} + \ri \mu \overline{\Phi} \partial_0 \hat{\phi}_2\,,  \nonumber\\
	\mathcal{L}_{\rm gauss} =& \frac{1}{2} \begin{pmatrix}
		\hat{\sigma} & \hat{\phi}_1 & \hat{\phi}_2 
	\end{pmatrix} 
	\begin{pmatrix}
		- \frac{1}{2g} &   \overline{\Phi} & 0 \\ 
		\overline{\Phi} & -  \Delta + m^2 +  \overline{\Sigma} - \mu^2 & 2 \ri \mu \partial_0 \\
		0 & -2 \ri\mu \partial_0 & - \Delta + m^2 + \overline{\Sigma} - \mu^2
	\end{pmatrix}
	\begin{pmatrix}
		\hat{\sigma} \\ \hat{\phi}_1 \\ \hat{\phi}_2 
	\end{pmatrix} \nonumber\\
&+ \frac{1}{2} \boldsymbol{\hat{\eta}} \left( -\Delta +m^2 +  \overline{\Sigma} \right) \boldsymbol{\hat{\eta}}\,,\nonumber\\
	\mathcal{L}_{\rm int} =& \frac{1}{2} \hat{\sigma} \left( \hat{\phi}_1^2 + \hat{\phi}_2^2 + \boldsymbol{\hat{\eta}}^2 \right) \,.
	\label{eq:O(N)_action_expanded}
\end{align}
Criticality is reached in this model by taking the formal limit $g \rightarrow \infty$ while tuning $m^2$ to a specific value depending on the regularization chosen and on the specific manifold the model \eqref{eq:O(N)_action} is considered. Computing the partition function with the action above as a saddle point expansion around a field saddle point configuration leads to the effective potential $V$ as follows
\begin{equation}
\mathcal{Z} = \int \mathcal{D} \boldsymbol{\phi} \mathcal{D} \sigma\, e^{- S[\boldsymbol{\phi}, \sigma]} = e^{- \text{Volume} \times V}, \qquad V= N V^{(-1)} + N^0 V^{(0)} + \dots\, .
\label{eq:V_general}
\end{equation}
Working on non-compact manifolds, the potential $V$ hereby defined is a potential density. When working on the cylinder, we consider the compact manifold $S^1_\beta \times S^{d-1}_R$ in the limit $\beta \rightarrow \infty$. In this case, we define the effective potential with appropriate (finite) volume factors as follows
\begin{equation}
\mathcal{Z} \sim e^{- \beta V_\text{cyl.}}\,, \quad \beta \rightarrow \infty\,.
\label{eq:V_cyl}
\end{equation}
Using the relations in appendix \ref{sec:termo}, we can relate the potential $V_\text{cyl.}$ to the energy $E(Q)$ of the saddle configuration, which is related to the scaling dimension $\Delta(Q)$ by the usual state-operator correspondence relation $\Delta(Q) = R E(Q)$.

\subsection{Review of the leading order result for the ground state energy.}

We start this section by reviewing the computation of the LO in $1/N$ result for the ground state energy at finite chemical potential on $\mathbb{R}^d$. At LO, the path integral defining the partition function with action expanded as in \eqref{eq:O(N)_action_expanded} is Gaussian and can be carried out exactly. The corresponding effective potential reads
\begin{align}
	V^{(-1)} &= \frac{1}{2} \Phi^2 (m^2-\mu^2+\Sigma) - \frac{\Sigma^2}{4g} + \underbrace{\frac{1}{2} \int \frac{\dd^d k}{(2\pi)^d} \log \left( k^2 + m^2 + \Sigma\right) }_{I}\,.
\end{align}
We specify the conformal point by taking $m^2 \rightarrow 0, \, g \rightarrow \infty$. The integral $I$ can be computed in various regularizations, for example in $\zeta$-function regularization one finds a finite result as long as $2<d<4$:
\begin{equation}
	\left[ I \right]_{\zeta,d} = - \frac{1}{2}\left. \frac{\dd}{\dd s} \right|_{s=0} \int \frac{\dd^d k}{(2\pi)^d} (k^2+\Sigma)^{-s}  = - \frac{\Gamma\left( - \frac{d}{2} \right)}{2(4\pi)^{\frac{d}{2}} } \Sigma^{\frac{d}{2}}  \xrightarrow[d=3]{}  - \frac{\Sigma^{\frac{3}{2}}}{12\pi}\,.
	\label{eq:LO_zeta}
\end{equation}
Alternatively, one can implement a cutoff $|k| < \Lambda$ and subtract the value of $I$ for $\mu = \Phi = \Sigma = 0$, obtaining\footnote{Throughout this section, we neglect the tails $\sim \Lambda^{-p}$ with $p>0$. In the NLO computation, we will also include them in order to numerically speed up the convergence for $\Lambda \rightarrow \infty$.}
\begin{equation}
 \left[	\mathcal{I} \right]_{\Lambda,d} \equiv \left[I- \left. I\right|_{\mu=0} \right]_{\Lambda,d} = \frac{2\Lambda^d}{ d^2 (4\pi)^{\frac{d}{2}} \Gamma\left( \frac{d}{2} \right)} - \frac{\Gamma\left( - \frac{d}{2} \right)}{2(4\pi)^{\frac{d}{2}}} \Sigma^{\frac{d}{2}} \xrightarrow[d=3]{} \frac{\Lambda^3}{18\pi^2} - \frac{\Sigma^{\frac{3}{2}}}{12\pi}\,.
\end{equation}
The $\zeta$-function result reproduces the finite part of the expression above (which is scheme-independent for $2<d<4$), and the divergence is canceled by a cosmological constant-type term. The structure of the divergence is sensible to the order of limits $d \rightarrow 3, \, \Lambda \rightarrow \infty$, if we were to reverse the order, one would obtain
\begin{equation}
	\left[	\mathcal{I} \right]_{\Lambda,d=3} =  \frac{\Lambda \Sigma}{4\pi^2} - \frac{\Sigma^{\frac{3}{2}}}{12\pi}\,.
	\label{eq:flat_we_use}
\end{equation}
In this case, the divergence is absorbed by tadpole terms, leaving the same scheme-independent finite part. At LO, the most straightforward way to compute such finite part is via $\zeta$-function regularization \eqref{eq:LO_zeta}, but, as it will be evident later, it becomes inapplicable at NLO, where one has to resort to the regularization \eqref{eq:flat_we_use}. 

The saddle equations and their non-trivial solutions are then found as
\begin{equation} \label{eq:sol_LOflat}
\begin{cases}	0 = \Phi_*(\Sigma_*  - \mu^2  ) \\
	0 = \Phi_*^2 - \frac{d \Gamma\left( - \frac{d}{2} \right)}{2(4\pi)^{\frac{d}{2}}} \Sigma_*^{\frac{d}{2}-1} 
\end{cases} \quad \implies \quad 	\Sigma_* = \mu^2 , \quad \Phi_*^2 = \frac{d}{2} \frac{\Gamma \left( - \frac{d}{2} \right)}{(4\pi)^{\frac{d}{2}}} \mu^{d-2}\,.
\end{equation}
Computing the potential at this saddle solution, one finds the energy and charge density of this configuration at leading order at large $N$ using appendix \ref{sec:termo}: 
%
\begin{equation}
\mathcal{V}^{(-1)}(\mu)\equiv \left. V^{(-1)}\right|_{\Sigma_*,\Phi_*}-\left. V^{(-1)}\right|_{\mu=0}=- \frac{\Gamma\left( - \frac{d}{2} \right)}{2(4\pi)^{\frac{d}{2}} } \mu^{d}\,, 
\end{equation}
\begin{align}
	\hat{\rho} &= \frac{d}{2} \frac{\Gamma\left( -\frac{d}{2} \right)}{(4\pi)^{\frac{d}{2}}} \mu^{d-1}\,, &&
	 \epsilon^{(-1)}(\hat{\rho}) = N\frac{d+1}{d} \left(- \frac{(4\pi)^{\frac{d}{2}}}{\Gamma\left(1-\frac{d}{2} \right)}  \right)^{\frac{1}{d-1}} \hat{\rho}^{\frac{d}{d-1}} \xrightarrow[d=3]{} \frac{4\sqrt{\pi}}{3} \hat{\rho}^{\frac{3}{2}}\,.
	 \label{eq:EOS_flat}
\end{align}
We recall that hatted quantities are normalized by $N$, i.e. $\hat{\rho} = \rho/N$. The computation on $S^1_\beta \times S^{d-1}_R$ follows similar steps, where in the action \eqref{eq:O(N)_action} we introduce a conformal coupling
\begin{equation}
	-\Delta_{\mathbb{R}^d} \rightarrow -\Delta_{S^1_\beta \times S^{d-1}_R} + \frac{(d-2)^2}{4 R^2}\, . 
\end{equation}
Separating out the conformal coupling term, criticality is again achieved at $m^2 \rightarrow 0, g \rightarrow \infty$ in mass-independent schemes. The resulting effective potential follows by replacing the integral $I$ by the sum\footnote{The prefactor difference comes from the definitions \eqref{eq:V_general} and \eqref{eq:V_cyl}.}
\begin{equation}
	I_{\text{cyl.}}=  \frac{1}{2 \beta} \sum_{n \in \mathbb{Z}} \sum_{\ell=0}^\infty \text{deg}(\ell) \log \left( \omega_n^2 + E_\ell^2 (\Sigma)\right)\,, 
\end{equation}
with
\begin{equation}
\text{deg}(\ell) = \frac{(d + 2 \ell - 2) \Gamma(d + \ell - 2)}{\Gamma(\ell + 1) \Gamma(d - 1)}, \qquad 
\omega_n = \frac{2\pi n}{\beta}\,,  \qquad E_\ell^2(\Sigma) = \frac{1}{R^2} \left( \ell + \frac{d-2}{2} \right)^2 + \Sigma\,.
\label{eq:cyl_notation}
\end{equation}
There are multiple ways to regulate this sum. For simplicity, we take from the start $d=3$ so that the Laplacian eigenvalue degeneracy is simply $2\ell+1$. In $\zeta$-function regularization, the result is straightforward \cite{Hellerman:2015nra, Monin:2016bwf}, here we focus on an analog of the regularization \eqref{eq:flat_we_use} which is more suited for the NLO computation and was introduced in a similar setting  \cite{Dyer:2015zha, Pufu:2013eda, Pufu:2013vpa, Chester:2021drl, Chester:2017vdh}. Taking the zero-temperature limit $\beta \rightarrow \infty$, we can introduce the combined cutoff $\omega^2 +\left( \ell + \frac{d-2}{2} \right)^2/R^2 < \Lambda^2$, so that the analog of \eqref{eq:flat_we_use} becomes\footnote{As it was implicitly done in the previous section, the regularization affects the sum/integral over frequencies. In flat space, this is a simple cutoff over momenta, while here it is a combined cutoff in the frequencies $\omega$ and the $\ell$ quantum number, hence the notation $[\dots]_\Lambda$ for the measure.}
\begin{equation}\label{eq:cyl_LO_reg}
\left[\mathcal{I}_{\text{cyl.}} \right]_{\Lambda,d=3} = \frac{1}{2} \left[ \sum_{\ell=0} \int \frac{\dd \omega}{2\pi} \right]_\Lambda (2\ell+1) \log \left(  1 + \frac{\Sigma}{\omega^2 + \frac{1}{R^2} \left( \ell + \frac{1}{2} \right)^2}  \right)\,.
\end{equation}
The expansion for $\Lambda \rightarrow \infty$ can be computed via the Euler-Maclaurin formula. The result comes automatically organized as an expansion for $\Sigma \rightarrow \infty$, where one finds
\begin{equation}
\left.[\mathcal{I}_{\text{cyl.}} \right]_{\Lambda,d=3}= \frac{R^2 \Lambda \Sigma}{\pi}  + \frac{1}{R} \left( - \frac{(\Sigma R^2)^{\frac{3}{2}}}{3} + \frac{(\Sigma R^2)^{\frac{1}{2}}}{24} -\frac{7}{1920 (\Sigma R^2)^{\frac{1}{2}}}  +\dots \right) \,.
\end{equation}
If we divide this quantity by the volume of $S^2_R$, its thermodynamical limit $R \rightarrow \infty$ precisely reproduces the result computed in $\mathbb{R}^3$ as expected:
\begin{equation}\label{eq:match_LO_flatcyl}
	\lim_{R \rightarrow \infty} \frac{\left[\mathcal{I}_{\text{cyl.}} \right]_{\Lambda,d=3}}{4\pi R^2}= \frac{\Lambda \Sigma}{4\pi^2} - \frac{\Sigma^{\frac{3}{2}}}{12 \pi}\,.
\end{equation}
The solution of the saddle for $\Sigma$ is now $\Sigma_{*\text{cyl.}}= \mu^2 - \frac{1}{4R^2}$ while $\Phi_{*\text{cyl.}}$ is now a non-trivial expansion in $\mu R$.  Removing the divergence, we get the potential at the minimum 
\begin{equation}
\mathcal{V}^{(-1)}_{\text{cyl.}} (\mu) = \frac{1}{R} \left\{  - \frac{(\mu R)^3}{3} + \frac{(\mu R)}{6} - \frac{1}{60 (\mu R)}   + \dots \right\}\,.
\end{equation}
From this expression, one can compute the total charge at LO on $S^2$ and invert its relation to express $\mu = \mu(Q)$,
\begin{equation} \label{eq:mu_hatQ}
	Q = - \frac{\partial}{\partial \mu}  N \mathcal{V}^{(-1)}_{\text{cyl.}} \quad \implies \quad \mu R = \hat{Q}^{\frac{1}{2}}+ \frac{1}{12}\hat{Q}^{-\frac{1}{2}} + \frac{7}{1440}\hat{Q}^{-\frac{3}{2}}  +O\left(\hat{Q}^{-\frac{5}{2}}\right) \,.
\end{equation}
This is useful in order to express the ground state energy $E(Q)$, analog of \eqref{eq:EOS_flat}, and the corresponding scaling dimension $\Delta_{-1}$ as
\begin{equation}\label{eq:Delta_LO}
 \Delta_{-1}(\hat{Q}) = \frac{2}{3} \hat{Q}^{\frac{3}{2}} + \frac{1}{6} \hat{Q}^{\frac{1}{2}} - \frac{7}{720} \hat{Q}^{-\frac{1}{2}} + O\left(\hat{Q}^{-\frac{3}{2}}\right) \,,
\end{equation}
Matching with the EFT prediction \eqref{eq:EFT_prediction} we can see that the $\beta_0, \beta_1$ coefficients do not enter at this order, but we only have contributions from the $\alpha_i$ coefficients, which stems from the ground-state-only contribution. These have an expansion of the form
\begin{equation} \label{eq:expansion_N_alpha}
\alpha_i=N^{-i}\left(N\alpha^{(-1)}_i+\alpha^{(0)}_i+O\left(N^{-1}\right)\right),
\end{equation}
from which we read $\alpha_{3/2}^{(-1)} = 2/3,\, \alpha_{1/2}^{(-1)} = 1/6, \, \alpha_{-1/2}^{(-1)} = -7/720$. These two coefficients can be directly related to the EFT coefficients $c_1, c_2,$ and $c_4$ using the results in appendix \ref{sec:EFT}, leading to
\begin{equation}
c_1 = \frac{N}{12 \pi} + O\left(N^0\right)\,, \qquad c_2 = \frac{N}{48 \pi} + O\left(N^0\right)\,,\qquad c_4 =- \frac{N}{960 \pi} + O\left(N^0\right)\,.
    \label{eq:Wilson_coeff_LO}
\end{equation}
In the next section, we will comment on the coefficient $c_3$ which does not enter into the EFT ground state contribution but can be matched from the GB dispersion relation. 

It is worthwhile to point out that the expansion (\ref{eq:Delta_LO}) for $\hat{Q} \ll 1$ perfectly matches the known orders of large-$N$ perturbation theory \cite{Alvarez-Gaume:2019biu,Giombi:2020enj}.

\section{The next-to-leading order computation}
\label{sec:NLO}
\subsection{Computation on the plane}

We start by computing the NLO effective potential in flat space. 
From the Lagrangian \eqref{eq:O(N)_action_expanded} one can read off the following critical propagators for the various fields
\begin{align}
	D_{\hat{\sigma}\hat{\sigma}} &= - \frac{1}{ \Phi^2} \left\{  \frac{4 \mu^2 k_0^2}{k^2+\Sigma-\mu^2} + \left(k^2+\Sigma-\mu^2\right)  \right\}, & &  D_{\hat{\sigma} \hat{\phi}_1} = D_{ \hat{\phi}_1 \hat{\sigma}} = \frac{1}{\Phi}\,,  \nonumber\\
	D_{\hat{\phi}_2 \hat{\phi}_2}&= \frac{1}{k^2 + \Sigma-\mu^2}, & & D_{\hat{\sigma} \hat{\phi}_2} = -  D_{ \hat{\phi}_2 \hat{\sigma}} = \frac{1}{\Phi} \frac{2 \mu k_0}{k^2+\Sigma-\mu^2}\,,  \nonumber\\
	D_{\hat{\eta}_i \hat{\eta}_j} &= \frac{\delta_{ij}}{k^2 + \Sigma}\,,  &&D_{\hat{\phi}_1 \hat{\phi}_1} = D_{\hat{\phi}_1 \hat{\phi}_2} = D_{\hat{\phi}_2 \hat{\phi}_1 } = 0\,.
	\label{eq:propagators_flat}
\end{align} 
As a preliminary result one needs the $\hat{\eta}_i$-bubble diagram, which amounts to the loop integral 
\begin{align}
	\Pi(p^2,m^2) &= \vcenter{\hbox{\includegraphics[scale=1.2]{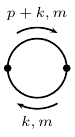}}} =  \frac{1}{2}\int \frac{\dd^d k}{(2\pi)^d} \frac{1}{[k^2+m^2][(k+p)^2+m^2]}  \, .
\end{align}
Solid lines correspond to $\hat{\eta}$-propagators from \eqref{eq:propagators_flat}. The bubble is convergent for $2<d<4$ and can be computed in closed form (see for example ref.~\cite{Marino:2021six}):
\begin{equation}
	\Pi(p^2,m^2) = m^{d-4} \frac{\Gamma\left( 2- \frac{d}{2} \right)}{2\pi (16\pi)^{\frac{d}{2}-1}} (x+4)^{\frac{d-4}{2}} {}_2 F_1 \left( 2-\frac{d}{2}, \frac{1}{2}, \frac{3}{2} , \frac{x}{x+4} \right), \quad x= p^2/m^2 \, .
	\label{eq:bubble}
\end{equation}
The bubble diagram is used to build the connected 2-point function of $\hat{\sigma}$ at leading order in $N$ in the unbroken phase $\mu = \Sigma =0$ as follows
\begin{align}
	\langle \hat{\sigma}(p) \hat{\sigma}(-p) \rangle &= \vcenter{\hbox{\includegraphics[scale=1.2]{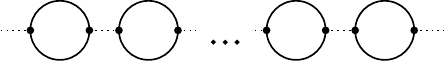}}} \nonumber\\
	&= (-2g) \sum_{j=0}^\infty \left(-2g\Pi(p^2,0)\right)^j  \nonumber\\ & \xrightarrow[g \rightarrow \infty]{} \quad  - \frac{1}{\Pi(p^2,0)} = 4^{d-1} \pi ^{\frac{d}{2}-\frac{3}{2}} \sin \left(\frac{\pi d}{2}\right) \Gamma \left(\frac{d-1}{2}\right)|p|^{4-d} \, ,
	\label{eq:Sigma_prop_flat}
\end{align}
%
where dashed lines correspond to $\hat{\sigma}$-propagators from \eqref{eq:propagators_flat}. The NLO contribution to the effective potential comprises of a leftover term from the LO $\hat{\eta}$-determinant (due to the fact that the $\hat{\eta}_i$ are $N-2$ in total): 
\begin{equation}
	V^{(0)}_{I} =-2 \times \frac{1}{2} \int \frac{\dd^d k}{(2\pi)^d} \log(k^2 +  \Sigma)\, ,
\end{equation}
the log-determinant of the three fields $\hat{\sigma}, \hat{\phi}_1, \hat{\phi}_2$ :
\begin{align}
	V^{(0)}_{II} &= \frac{1}{2} \int \frac{\dd^d k}{(2\pi)^d} \log \det \begin{pmatrix}
		0 & \Phi & 0 \\\Phi & k^2 + \Sigma- \mu^2 & -2 \mu k_0 \\ 0 & 2\mu k_0 & k^2 + \Sigma - \mu^2
	\end{pmatrix} \nonumber\\
&= \frac{1}{2} \int \frac{\dd^d k}{(2\pi)^d} \log \{ - \Phi^2 \left( k^2 + \Sigma - \mu^2 \right) \} \, ,
\end{align}
and a ring diagram obtained from the resummed propagator \eqref{eq:Sigma_prop_flat} with $\mu , \Sigma \neq 0$:
\begin{align}
	V^{(0)}_{III} &= \vcenter{\hbox{\includegraphics[scale=1.2,angle=72]{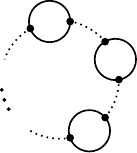}}}= - \sum_{j=1}^\infty \frac{1}{2j} \int \frac{\dd^d k}{(2\pi)^d} \left[ - \frac{1}{\Phi^2}  \left(  \frac{4\mu^2 k_0^2}{k^2+\Sigma - \mu^2} + k^2  + \Sigma - \mu^2 \right) \Pi(k^2, \Sigma) \right]^j  \nonumber\\
	&= \frac{1}{2} \int \frac{\dd^d k}{(2\pi)^d} \log \left\{  1 + \frac{1}{\Phi^2}  \left(  \frac{4\mu^2 k_0^2}{k^2+\Sigma - \mu^2} + k^2  + \Sigma - \mu^2 \right) \Pi(k^2, \Sigma) \right\} \, .
\end{align}
Alternatively, one can integrate out the $\boldsymbol{\hat{\eta}}$ fluctuations to obtain an effective action for $\hat{\sigma}$, expand it to the quadratic order and compute the fluctuation determinant. It is easy to show that such computation includes the last two pieces above:
\begin{equation}
V^{(0)}_{II}  + V^{(0)}_{III}  = \frac{1}{2} \int \frac{\dd^d k}{(2\pi)^d} \log \det \begin{pmatrix}
		- \Pi(k^2,\Sigma) & \Phi & 0 \\\Phi & k^2 + \Sigma- \mu^2 & -2 \mu k_0 \\ 0 & 2\mu k_0 & k^2 + \Sigma - \mu^2
	\end{pmatrix} \, .
\label{eq:effective_action_NLO}
\end{equation}
Putting all together, the (subtracted) effective NLO potential at the saddle (\ref{eq:sol_LOflat}) reads
\begin{align}
\mathcal{V}^{(0)}  &= \frac{1}{2} \int \frac{\dd^d k}{(2\pi)^d} \log \left(\frac{k^2\Phi^2_*+\left(4 \mu^2 k_0^2+k^4\right)\Pi(k^2,\Sigma_*)}{k^4\Pi(k^2,0)} \right) - \int \frac{\dd^d k}{(2\pi)^d} \log \left( 1 + \frac{\Sigma_*}{k^2} \right)\,.
\end{align}
Let's compute it in $d=3$. We regularize it implementing a cutoff $\Lambda$ as done at LO in eq. ~(\ref{eq:flat_we_use}), we find 
\begin{align}\label{eq:NLO_Vflat_result}
\left[\mathcal{V}^{(0)}\right]_{\Lambda,d=3}  &= \frac{1}{4\pi}\left(-\frac{2}{3\pi}\Lambda \mu^2  + v^{(0)} \mu^3\right) \,\ \ \qquad \text{with} \qquad v^{(0)}=-0.03530346\dots\,.
\end{align}
The quantity $v^{(0)}$ is given in term of a definite integral that has no simple relations to known functions but which can be evaluated with arbitrarily high numerical precision.
Expanding the Legendre transform at NLO at large $N$, see appendix \ref{sec:termo}, it is straightforward to compute the next correction to the energy density of the ground state:
\begin{equation}\label{eq:NLO_flat_result}
\epsilon(\hat{\rho}) = \left(  \frac{4 \sqrt{\pi}}{3} + 2\sqrt{\pi}v^{(0)} \frac{1}{N} + O\left(N^{-2}\right)\right) \hat{\rho}^{\frac{3}{2}}\,.
\end{equation}

\subsection{Dispersion relation for the conformal Goldstone}

Before moving to the computation of $\Delta_0$, let us extract further information from the effective large-$N$ potential \eqref{eq:effective_action_NLO}. In flat space, for large charge density $\rho$, it predicts the existence of a mode with the following (Lorentzian) dispersion relation:
\begin{align}
	\omega^2 &=  \frac{p^2}{2} + \frac{p^4}{96\pi \hat{\rho}} + O\left(\frac{p^6}{\hat{\rho}^2}\right)\,,
\end{align}
which can be identified with the universal conformal GB appearing in the EFT \eqref{eq:EFT}. Within the EFT, the dispersion relation of the Goldstone mode is found to be \cite{Cuomo:2020rgt}
\begin{equation}
	\omega_\ell = \frac{1}{\sqrt{d-1}} \lambda_\ell + \frac{\gamma}{Q^{\frac{2}{d-1}}} \left( \frac{\lambda_\ell^3 R^2}{d-1} - \lambda_\ell \right) + \mathcal{O}\left( \frac{\lambda_\ell^5 R^4}{Q^{\frac{4}{d-1}}} \right) ,\quad \lambda_\ell = \frac{1}{R} \sqrt{\ell(\ell+d-2)}  \, .
\end{equation}
The coefficient $\gamma$ can also be measured in flat space if one takes the formal macroscopic limit as follows
\begin{equation}
	E,Q,\ell \rightarrow \infty
	\quad \text{with} \quad \omega_\ell \rightarrow \omega, \quad \lambda_\ell \rightarrow |p|, \quad \frac{Q}{\text{Vol}(S^{d-1})} \rightarrow \rho, \quad \omega,p^2,\rho \,\, \text{fixed.}
\end{equation}
Restricting to $d=3$, the EFT predicts the following form:
\begin{equation}
	\omega = \frac{|p|}{\sqrt{2}} + \frac{ \gamma}{8\pi N \hat{\rho}}  |p|^3 + \mathcal{O}\left( \frac{|p|^5}{16\pi^2 N\hat{\rho}^2} \right)\,.
\end{equation}
Comparing with our large-$N$, result one finds the following estimate for the $\gamma$-parameter:
\begin{equation}
	\gamma = \frac{N}{12\sqrt{2}} + O\left(N^0\right)\,,
\end{equation}
the parameter $\gamma$ is not an independent piece of data, it is related to the Wilson coefficients $c_1, c_2, $ and $c_3$ as follows
\begin{equation}
	\gamma = \frac{[c_2(d-2)+c_3](d-2)}{c_1^{\frac{d-3}{d-2}}d \sqrt{d-1} (d \Omega_{d-1})^{- \frac{2}{d-1}}} \xrightarrow[d \rightarrow 3]{} \sqrt{8\pi^2} \left(c_2+c_3\right)\, .
	\label{eq:gamma}
\end{equation}
Inputting the value of $c_2$ at leading order found in \eqref{eq:Wilson_coeff_LO}, we see that the coefficient $c_3$ vanishes at leading order, namely $c_3 = O(N^0)$.

The coefficient $c_3$ turns out to be zero also at leading order in the $\epsilon$-expansion for the $\phi^6$ tricritical model in $d=3-\epsilon$ \cite{Badel:2019khk}. From the point of view of the EFT, the operator corresponding to $c_3$ does not violate any (accidental) symmetry that we can find, so there is no a priori reason why this should vanish at all orders.

\subsection{Computation on the cylinder}

The computation on the cylinder follows similar steps as the flat space one. All fields in the Lagrangian are decomposed in Fourier modes as follows
\begin{equation}
	f = \frac{1}{\text{Vol}(S^1_\beta \times S^2_R)^{\frac{1}{2}}} \sum_{n \in \mathbb{Z}} \sum_{\ell \geq0} \sum_{m=-\ell}^\ell e^{i \omega_n \tau} Y_{\ell m}(\hat{n}) f_{n\ell m}, \quad f_{n,\ell, m}^* = f_{-n,\ell,-m},
\end{equation}
where in intermediate steps it is convenient to consider the ``thermal" geometry $S^1_\beta \times S^2_R$ and send $\beta \rightarrow \infty$ only at the end, $\omega_n$ are the corresponding bosonic Matsubara frequencies. The propagators can be read off the quadratic action
\begin{align}
	S_{\rm gauss} &= \frac{1}{2} \sum_{n \ell m} \begin{pmatrix}
		\hat{\sigma}_{n \ell m} & \hat{\phi}_{1, n \ell m} & \hat{\phi}_{2,n \ell m} 
	\end{pmatrix} 
	\begin{pmatrix}
		- \frac{1}{2g} &   \Phi & 0 \\ 
		\Phi &  \omega_n^2 + E_\ell^2(\Sigma)- \mu^2 & -2 \mu\omega_n\\
		0 & 2\mu \omega_n &  \omega_n^2 + E_\ell^2(\Sigma) - \mu^2
	\end{pmatrix}
	\begin{pmatrix}
		\hat{\sigma}_{n \ell m}^* \\ \hat{\phi}^{*}_{1, n \ell m} \\ \hat{\phi}^{*}_{1, n \ell m} 
	\end{pmatrix} \nonumber\\
& +\frac{1}{2} \sum_{n \in \mathbb{Z}} \sum_{\ell ,m } \left\{ \omega_n^2 + E_\ell^2(\Sigma) \right\} |\boldsymbol{\hat{\eta}}_{n \ell m}|^2\,,
\end{align}
where $E_\ell(\Sigma)$ is the eigenvalue of the sphere conformal Laplacian with a mass term $\Sigma$,\footnote{In order to make the notation less heavy, often, when clear, we will avoid writing the explicit dependence on $\Sigma$.} already introduced in eq.~(\ref{eq:cyl_notation}). The $\hat{\sigma} \boldsymbol{\hat{\eta}} \boldsymbol{\hat{\eta}}$ vertex action instead is
\begin{equation}
	S_{\rm int} = \frac{1}{2 \sqrt{4\pi R^2 \beta}} \sum_{n _i} \sum_{\ell_i m_i} \delta_{n_1, - n_2 - n_3} \langle \ell_1 m_1 | \ell_2 m_2 ; \ell_3 m_3 \rangle \hat{\sigma}_{n_1 \ell_1 m_1} \boldsymbol{\hat{\eta}}_{n_2 \ell_2 m_2} \boldsymbol{\hat{\eta}}_{n_3 \ell_3 m_3} 
\end{equation}
where the symbol $\langle \ell_1 m_1 | \ell_2 m_2 ; \ell_3 m_3 \rangle$ stems for the triple integral of spherical harmonics, see appendix \ref{sec:AppSphHarm}.  
The bubble amplitude analog to the flat space result \eqref{eq:bubble} is
\begin{align}
	\Pi_{n \ell m }^{n' \ell' m'} 
	&= \frac{1}{8 \pi R^2 \beta} \delta_{n n'} \sum_{n_1} \sum_{\ell_1 m_1} \sum_{\ell_2 m_2}  \frac{  \langle \ell m | \ell_1 m_1 ; \ell_2 m_2 \rangle   \langle \ell' m' | \ell_1 m_1 ; \ell_2 m_2 \rangle }{\left[\omega_{n_1}^2 + E_{\ell_1}^2(\Sigma) \right] \left[\omega_{n+n_1}^2 + E_{\ell_2}^2(\Sigma)\right]} \nonumber\\
	&=  \delta_{n n'} \delta_{\ell \ell'} \delta_{m m'} \frac{1}{16 \pi R^2} \sum_{\ell_1\ell_2}  (2\ell_1+1)(2\ell_2+1) \begin{pmatrix}
		\ell_1 & \ell_2 & \ell \\ 0 & 0 & 0
	\end{pmatrix}^2 \frac{E_{\ell_1}+E_{\ell_2}}{E_{\ell_1}E_{\ell_2}} \frac{1}{\omega_n^2 + [E_{\ell_1} + E_{\ell_2}]^2 } \nonumber\\
	&=  \delta_{n n'} \delta_{\ell \ell'} \delta_{m m'} \Pi_{n \ell}(\Sigma)\,.
\end{align}
Correspondingly,  at large $N$ the resummed propagator for $\hat{\sigma}$ reads, in Fourier space,
\begin{equation}
	\langle \hat{\sigma}_{n \ell m} \hat{\sigma}_{n \ell m} \rangle  = (-2g) \sum_{j=0}^\infty \left(-2 g \Pi_{n \ell}(\Sigma)\right)^j = \frac{-2g}{1+2g \Pi_{n \ell}(\Sigma)} \xrightarrow[g \rightarrow \infty]{} - \frac{1}{\Pi_{n \ell}(\Sigma)} \,.
\end{equation}
As in the case of flat space, the contributions to the NLO of the effective potential coming from the bubble of $\hat{\sigma}-\hat{\phi}_1-\hat{\phi}_2$ propagator combines to
\begin{equation}
V^{(0)}_{II, \text{cyl.}}  + V^{(0)}_{III, \text{cyl.}}  = \frac{1}{2\beta} \sum_{n \in \mathbb{Z}} \sum_{\ell=0}^\infty (2\ell+1) \log \det \begin{pmatrix}
		- \Pi_{n\ell} & \Phi & 0 \\ \Phi & \omega_n^2 + E_\ell^2- \mu^2 & -2 \mu \omega_n \\ 0 & 2\mu \omega_n & \omega_n^2 + E_\ell^2 - \mu^2
	\end{pmatrix} \,.
\end{equation}
Following similar steps as before, i.e. adding the leftover term from the $\hat{\eta}$-determinant and substituting the VEVs with their values at the saddle, we find
\begin{align} \label{eq:NLO_cyl}
	\mathcal{V}^{(0)}_{\text{cyl.}}(\mu)&= \frac{1}{2\beta} \sum_{n \ell} (2\ell+1) \nonumber \\ &\left(\log \left(\frac{\left(\omega_n^2+\frac{\ell(\ell+1)}{R^2}\right) \Phi^2_{*\text{cyl.}}+\left(4 \mu^2 \omega_n^2+\left(\omega_n^2+\frac{\ell(\ell+1)}{R^2}\right)^2 \right)\Pi_{n \ell}\left(\mu ^2-\frac{1}{4R^2}\right)}{\left(\omega_n^2+\frac{(\ell+1/2)^2}{R^2}\right)^2 \Pi_{n \ell}(0)} \right) \right.\nonumber \\
	& \left.- 2 \log \left( 1 + \frac{\mu ^2-\frac{1}{4R^2}}{\omega_n^2+\frac{(\ell+1/2)^2}{R^2}} \right) \right) \nonumber \\
	&\eqqcolon \frac{1}{\beta}\sum_{n \ell} \left(\ell+\frac{1}{2}\right) f(\omega_n,\ell;\mu)\,.
\end{align}
To compute it, once again we take the zero-temperature limit $\beta \to \infty$ and we introduce the combined hard cutoff of eq.~(\ref{eq:cyl_LO_reg}):
\begin{equation}
\left[	\mathcal{V}^{(0)}_{\text{cyl.}}\right]_{\Lambda,d=3} = \left[ \sum_{\ell=0} \int \frac{\dd \omega}{2\pi} \right]_\Lambda \left(\ell+\frac{1}{2}\right) f(\omega,\ell;\mu)\,.
\end{equation}
This sum and integral are too complicated to perform analytically. We have the integral over $\omega$ and the sums,  the one over $\ell$ and the other two inside the function $\Pi$, over $\ell_1$ and $\ell_2$. Thus, we will resort to a numeric approach.

\subsection{Numerical implementation}
\label{subsec:Numercal}

Let us begin by looking at each of the sums in more detail, similarly to what was done in ref.~\cite{Pufu:2013eda},  where an analogous study has been performed in a different theory.  Let's look at $\Pi_{\ell}( \omega, \Sigma)$,\footnote{The use the notation $\Pi_{\ell}( \omega_n, \Sigma)\coloneqq\Pi_{n\ell}(\Sigma)$. 
} for fixed $\ell$ and $\ell_1$ the sum over $\ell_2$ has finitely many terms since the 3-$j$ symbol is zero unless the triangular equality is satisfied. The remaining sum over $\ell_1$ is convergent.  Moreover, it is not difficult to write down the large-$\ell_1$ asymptotics as a function of $\ell,\, \omega$,  and $\Sigma$.  We have
\begin{equation}\label{eq:exp_ell1}
\frac{1}{16\pi R}\left(\frac{1}{\ell_1^2}-\frac{1}{\ell_1^3}-\frac{3+6\Sigma R^{2}+2\ell(\ell+1)-\omega^2R^{2}}{4\ell_1^4}+O\left(\ell_1^{-5}\right)\right)\,.
\end{equation}
We can exploit this expansion to achieve great accuracy.  
Indeed, we perform the sum over $\ell_1$ explicitly up to a value $\overline{\ell_1}$ and then perform analytically the sum of the asymptotic expansion (\ref{eq:exp_ell1}),  in our computation with an approximation up to $\ell_1^{-20}$,  from $\overline{\ell_1}$ up to infinity.
%
For the special case $\Pi_{\ell}( \omega, 0)$, the double sum can be performed analytically and gives
\begin{equation}
\Pi_{\ell}( \omega, 0) = \frac{1}{32 R}\left|\frac{\Gamma\left(\frac{\ell+1+\ri \omega}{2}\right)}{\Gamma\left(\frac{\ell+2+\ri \omega}{2}\right)}\right|^2\,.
\end{equation}
Let us focus now on the integration over $\omega$ and the sum over $\ell$. We expect the result to be divergent, more precisely, linearly divergent in $\Lambda$ as in the flat-space computation. Similarly to what was just done for $\ell_1$,  an asymptotic expansion at large $\ell$ and $\omega$ will allow us not only to catch the divergence but also to improve our final accuracy thanks to analytic inputs.
The most challenging part is to compute the expansion of $\Pi_{\ell}( \omega, \mu^2-\frac{1}{4})$. We followed an approach very similar to the one used in the appendix C of ref.~\cite{Dyer:2015zha}. In our appendix \ref{sec:AppAsymp}, we give an expression of $\Pi_\ell$ we obtained.  Once the asymptotic expansion of the bubble is computed (see eqs.~\ref{eq:Pi_exp_largelomega} and \ref{eq:Pi0_exp_largelomega}) it's enough to insert it in eq.~(\ref{eq:NLO_cyl}).  We find at large $\ell$ and $\omega$\footnote{To lighten the notation we used the $l\coloneqq \left(\ell+\frac{1}{2}\right)/R$ introduced in appendix \ref{sec:AppAsymp}.}
\begin{align}
f(\omega_n,\ell;\mu) &= 2\mu^2\frac{-l^2 + \omega^2}{\left(l^2+\omega^2\right)^2}+ \frac{8\Delta_{-1}(\mu)}{\pi R^3}\frac{l^2 - 2 \omega^2}{\left(l^2+\omega^2\right)^{\frac{5}{2}}} \nonumber \\
&+\frac{\mu^2}{2R^2} \frac{(2\left(R\mu\right)^2-1)l^4 +(2\left(R\mu\right)^2+1) l^2\omega^2+(-14\left(R\mu\right)^2+3) \omega^4}{\left(l^2+\omega^2\right)^4} \nonumber \\
& + O\left(\left(l^2+\omega^2\right)^{-\frac{5}{2}}\right)\,,
\end{align}
where $\Delta_{-1}(\mu) \equiv \Delta_{-1}(\hat{Q}(\mu))$. 
The first term is responsible for a linear divergence in the cutoff, i.e.
\begin{equation}
2 \left(R\mu\right)^2 \left[ \sum_{l=1/(2R)} \int \frac{\dd \omega}{2\pi} \right]_\Lambda l \frac{-l^2 + \omega^2}{\left(l^2+\omega^2\right)^2}=\frac{1}{R}\left(-\frac{2}{3\pi}R^3\Lambda\mu^2+O\left((R\Lambda)^{-\frac{1}{2}}\right)\right) \,.
\end{equation}
As expected, dividing by the volume of $S^2_R$, the divergent piece matches the one found in the flat space computation (\ref{eq:NLO_Vflat_result}), where we adopted the same kind of regularization. This is akin to what shown in eq.~(\ref{eq:match_LO_flatcyl}) at LO.
The asymptotic expansion at large $\ell$ and $\omega$ provides us with much more useful information. Indeed, integrating it term by term we can write down a whole series, past the leading term, around $\Lambda \to \infty$ for our regularized effective potential. We get 
\begin{align} \label{eq:tails}
\left[	\mathcal{V}^{(0)}_{\text{cyl.}}\right]_{\Lambda,d=3} = & \frac{1}{R} \left(-\frac{2}{3\pi}R^3\Lambda\mu^2+R\mathcal{V}^{(0)}_{\text{cyl.}}(\mu)+g_{\frac{1}{2}}\frac{(R\mu)^2}{\sqrt{R\Lambda}} + \frac{(R\mu)^2 \left(12 (R\mu)^2-5\right)}{20 \pi R\Lambda}\right. \nonumber \\
& \left.  +\frac{g_{\frac{3}{2}}(R\mu)^2-\frac{4}{\pi}g_{\frac{1}{2}}\Delta_{-1}(\mu)}{(R\Lambda)^{\frac{3}{2}}} +O\left((R\Lambda)^{-\frac{5}{2}}\right)\right)\, ,
\end{align}
where we introduced the coefficients $g_{\frac{1}{2}}=0.187163\dots$ and $g_{\frac{3}{2}}=-0.0286809\dots$ which we can compute up to arbitrarily high numerical precision.    Notice that the expansion has no term proportional to $\Lambda^{-2}$.
We are ultimately interested in the finite result when we take the cutoff to infinity. We extract this quantity by computing the regulated function $\left[\mathcal{V}^{(0)}_{\text{cyl.}}\right]_{\Lambda,3}$ from many values of $\Lambda$ and extrapolating for $\Lambda \rightarrow \infty$.
We can use the knowledge of coefficients of the tails, i.e.  the terms that go like $\Lambda^{-n/2}$, to greatly improve the convergence for $\Lambda\to\infty$, allowing us to achieve more accurate final results with the same computational effort.  In our computation, we calculated the coefficients of the terms up to $\Lambda^{-3}$ with arbitrary precision. See the left panel of figure \ref{fig:Lambda_smallQ}, where, as an example, the regulated effective potential is plotted for $\mu =20$,\footnote{In the discussion that follows we take $R=1$ out of notational convenience, so that $\mu, \Lambda$ are dimensionless.} and the improvement in the convergence is evident.
Concretely, to extrapolate the renormalized effective potential $\mathcal{V}^{(0)}_{\text{cyl.}}(\mu)$ we compute the regulated quantity for different values of the cutoff,  subtract the calculated terms in the expansion at $\Lambda\to\infty$ and then fit to the results a function of this type
\begin{equation}
w^{[k]}(\mu)=w_0^{[k]}(\mu)+\sum_{i=i_0}^{k} w_i(\mu)\Lambda^{-\frac{i}{2}} \qquad \text{with} \quad i_0=7
\end{equation}
and read off the first term. Clearly, we have to take a value for $k$ comparable to the number of points that we are fitting to avoid overfitting. We estimated the final uncertainty comparing the results for fits with different values of $k$ and different numbers of points. The final computational effort grows substantially with $\mu$ since, to achieve a comparable accuracy,  a larger intermediate cutoff $\overline{\ell_1}$ and more points in $\Lambda$ are required.
For example, for $\mu =20$ we computed 24 points in $\Lambda$,  in the range $\Lambda \in [10,80]$. We used intermediate cutoff $\overline{\ell_1}=100$ for $\Lambda \leq 34$ and $\overline{\ell_1}=100+(34 -\Lambda)5$ for $\Lambda > 34$,  by doing so all uncertainties were smaller than $10^{-17}$.  For the fit, we looked at the values $h_0^{[k]}(20)$ with $k \in [10,25]$,  found a region of stability around $k \in [13,16]$ from which we extracted our final estimation with its uncertainty.  It turned out to be of order $10^{-9}$. For this computation done in Mathematica, the total single-core time required at 2.3GHz was roughly 24 hours.

\subsection{Final result and discussion}

\begin{figure}[t!]
\centering
  \raisebox{0\height}{\includegraphics[width=0.53\textwidth]{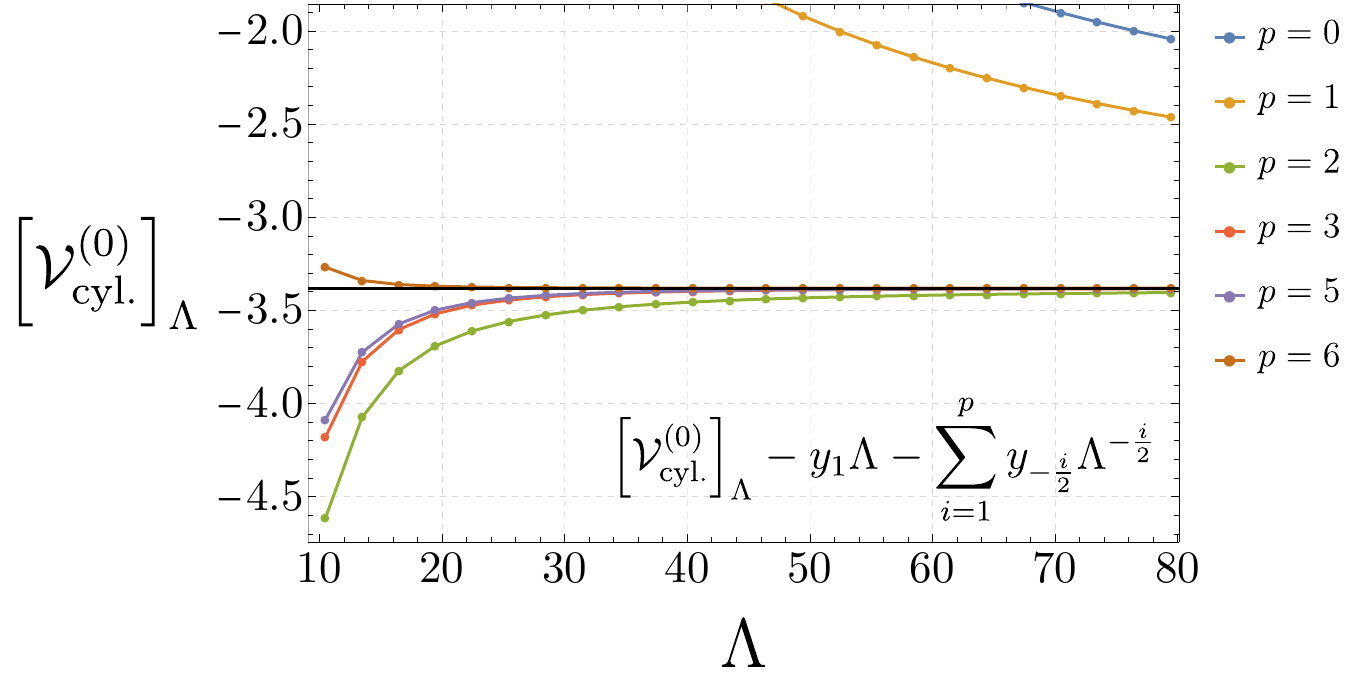}}
  \hspace*{.15cm}
  \raisebox{-0.035\height}{\includegraphics[width=0.44\textwidth]{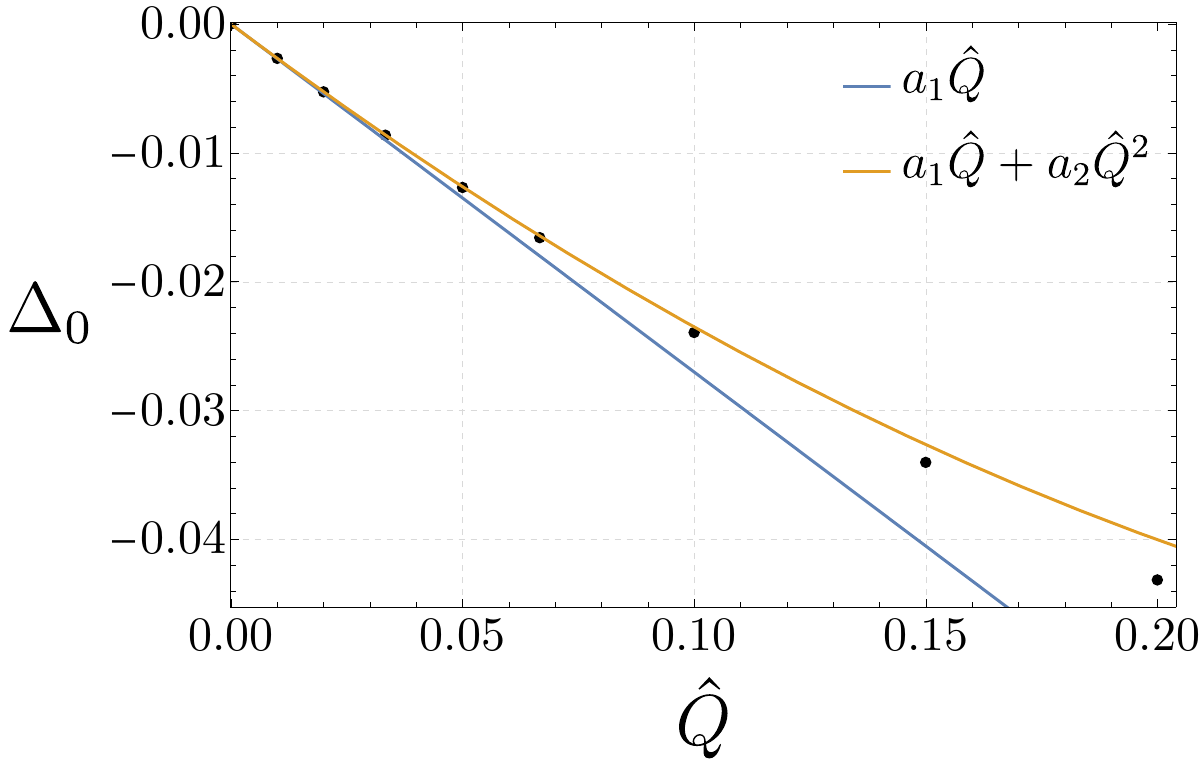}}
  \caption{\emph{Left panel}: The regulated effective potential on the cylinder at NLO at large $N$ for $\mu=20$ is plotted as a function of the cutoff $\Lambda$ after subtracting the linear divergence.  The index $p$ indicates up to which term of the form $\Lambda^{-p/2}$ we are subtracting, as written in the expression in the bottom part of the plot. The labels $y_i$ indicate the coefficients of eq.~(\ref{eq:tails}) for $\mu=20$, the black line the renormalized effective potential $\mathcal{V}^{(0)}_{\text{cyl.}}(\mu=20)$. Notice that we didn't plot the curve for $p=4$ since it would be the same as $p=3$ given that $y_{-2}=0$ (\ref{eq:tails}). \emph{Right panel}: Comparison between some values $\Delta_0(\hat{Q})$ in the regime $\hat{Q} \ll 1$ and the first two contributions of its small-$\hat{Q}$ expansion whose coefficients are known from perturbation theory (\ref{eq:coeff_smallQ_pt}).}
  \label{fig:Lambda_smallQ}
\end{figure}

In the previous section, we described how we can evaluate numerically the NLO of the effective potential at large $N$ on the cylinder $\mathbb{R} \times S^2_R$ at a given chemical potential $\mu$. Using the results in appendix \ref{sec:termo}, this can ultimately related to the function $\Delta_0(\hat{Q})$ introduced in \eqref{eq:LargeN_Delta}. In this section we present our results for the regimes $\hat{Q} \ll 1$ and $\hat{Q} \gg 1$.  

Let us start from the small-charge regime in which $\hat{Q} \ll 1$. In this regime, we expect an expansion of the scaling dimension of $\Delta_0(\hat{Q})$ in powers of $\hat{Q}$, starting from a linear term. Namely,
\begin{equation} \label{eq:smallQ_Delta0}
\Delta_0(\hat{Q})= a_1 \hat{Q} + a_2 \hat{Q}^2 + O\left(\hat{Q}^3\right)\,.
\end{equation}
The values of the first two coefficients are known from computations in large-$N$ perturbation theory.
They can be extracted from the anomalous dimension of the operator of charge $Q$ at order $1/N$ \cite{Lang:1992zw} and $1/N^2$ \citep{Derkachov:1997ch}.\footnote{As mentioned in the footnote 8 of ref.~\cite{Giombi:2020enj}, there is a typo in eq.~(5.23) of ref.~\cite{Derkachov:1997ch}: there is a missing factor $(d/2-1)^2$ in the first term in the bracket.} Their values are
\begin{equation} \label{eq:coeff_smallQ_pt}
 a_1= -\frac{8}{3\pi^2}\,, \qquad \qquad a_2=  - \frac{16(\pi^2-12)}{\pi^4}\,.
\end{equation}
In order to study the small-charge regime, we computed $\Delta_0(\hat{Q})$ for many values of $\hat{Q} \in [0, \frac{1}{2}]$.  In figure \ref{fig:Lambda_smallQ}, right panel, we plotted these results for the NLO of the scaling dimension together with the behavior predicted by the small-$Q$ expansion with the coefficients of eq.~(\ref{eq:coeff_smallQ_pt}).  We checked that, as expected, the best fit for our points is a polynomial fit that starts from a linear contribution.
Extracting from the fit numerical estimates for the coefficients, we find an excellent agreement with many significant digits with the values coming from perturbation theory, see table \ref{tab:smallQ}.  This serves as a first nontrivial check of our computation.   
\begin{table}[t!]
  \centering
  \begin{tabular}{c||c|c| c | c}
    \hline
    		& $a_1$	& $a_2$		& $a_3$		& $a_4$	 \\    \hline
    Perturbation theory 	& $-0.27018982304\dots$		& $0.34992965\dots$		& 	& 	 \\    \hline
    Fit (this work) 	& $-0.2701898230(1)\ \ $		& $0.34992968(8)\ $		& $-0.474333(4)$	& $0.4845(9)$	\\    \hline
  \end{tabular}
  \caption{Comparison between the coefficients of the small-charge expansion of the scaling dimension of the heavy operator with charge $Q$ at NLO at large $N$, see eq.~(\ref{eq:smallQ_Delta0}). In the first line are reported the numerical values for the coefficients $a_1$ and $a_2$ that have been computed via perturbation theory in refs.~ \cite{Lang:1992zw} and \cite{Derkachov:1997ch}, respectively. In the second one our results, extracted by fitting. In order to maximize their accuracy,  the estimates for $a_3$ and $a_4$ have been computed after inputting in the fit the analytical values for $a_1$ and $a_2$, given the excellent agreement of the latter with their theoretical values.}
  \label{tab:smallQ}
\end{table}
The small-$\hat{Q} $ expansion of $\Delta$ contains an infinite number of terms from the point of view of the $1/N$ expansion.  For $\Delta_0$, that we are studying here, those with the second highest power of $Q$ at each order in $1/N$.  From our numerical computation, it is possible to go beyond the first two.
In fact, now that we are confident in the coefficients of the first two terms, we can input their analytical values in the fit and extract the next coefficients more precisely.  In table \ref{tab:smallQ} we report the $a_3$ and $a_4$.\footnote{
The analytic from of these coefficients seems to be $a_k=\sum_{i=1}^k b_i\pi^{-2i}$ with $b_i$ rational numbers. Modulo some extra assumptions on the $b_i$, it would be possible to extract the analytical expression for,  e.g., $a_3$. With the level of accuracy reached, we still have several candidates of this form.  With more computational time, it would be possible to single out the exact one.}

Let's now move to the large-$\hat{Q}$ regime, namely $\hat{Q} \gg 1$. In this regime, the expected expansion of the NLO of the scaling dimension is
\begin{equation} \label{eq:largeQ_NLO}
\Delta_{0}(\hat{Q}) = \alpha^{(0)}_{\frac{3}{2}} \hat{Q}^{\frac{3}{2}} + \alpha^{(0)}_{\frac{1}{2}} \hat{Q}^{\frac{1}{2}} + \beta_0 + \alpha^{(0)}_{-\frac{1}{2}} \hat{Q}^{-\frac{1}{2}} + \beta^{(0)}_{-1} \hat{Q}^{-1} + O\left(\hat{Q}^{-\frac{3}{2}}\right)\,.
\end{equation}
It is noteworthy that at this order appear, together with the terms coming from the ground state configuration of the EFT (\ref{eq:GS_contribution}), also terms due to contributions of quantum corrections (\ref{eq:EFT_prediction}), absent in the LO (\ref{eq:Delta_LO}). To study this expansion starting from our computation, once again, we evaluated $\Delta_{0}(\hat{Q})$ for many values of $\hat{Q}$ and looked at its behavior. In the interval $\hat{Q} \in [1,50]$, we have computed 33 points. First of all, we checked which is the leading behavior in $\hat{Q}$; it turns out to be $\hat{Q}^{3/2}$, as expected. Then, we investigated the subleading contributions and find that the best fit is 
\begin{figure}[t!]
\centering
  \raisebox{0\height}{\includegraphics[width=0.47\textwidth]{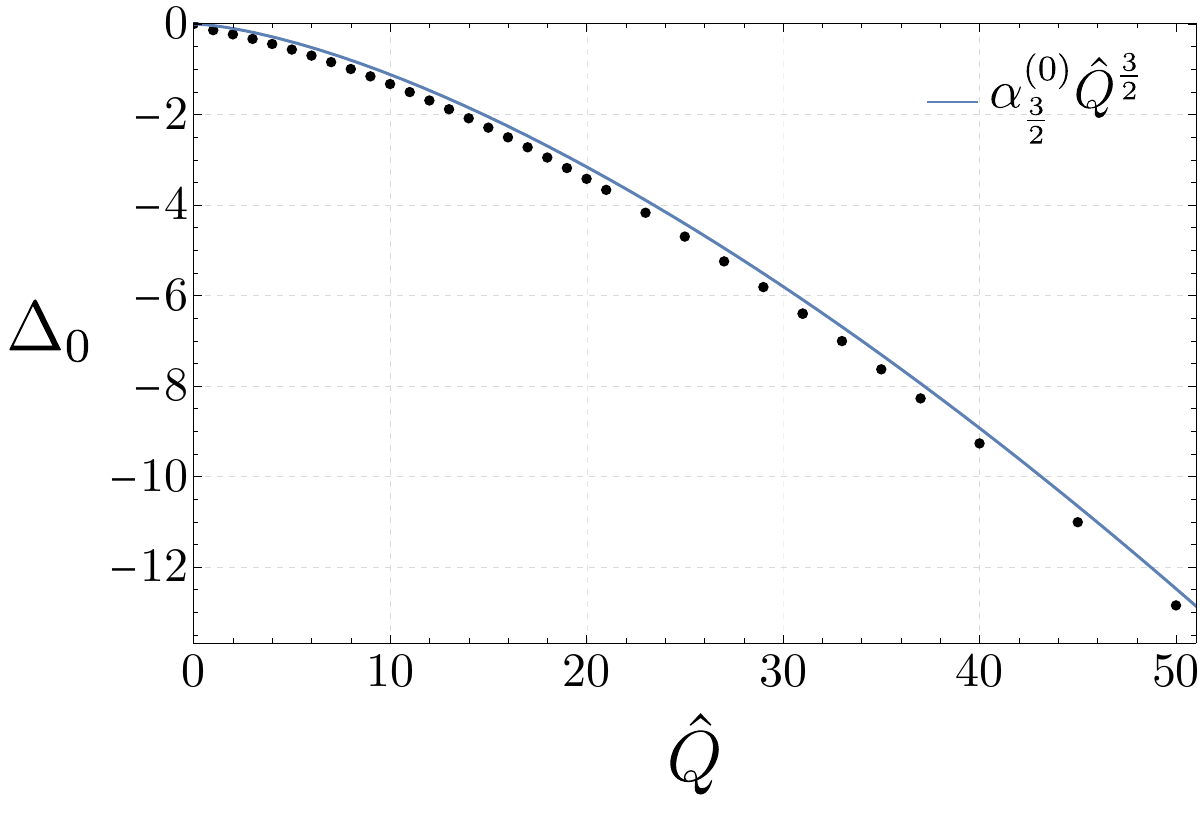}}
  \hspace*{.1cm}
  \raisebox{-0.02\height}{\includegraphics[width=0.505\textwidth]{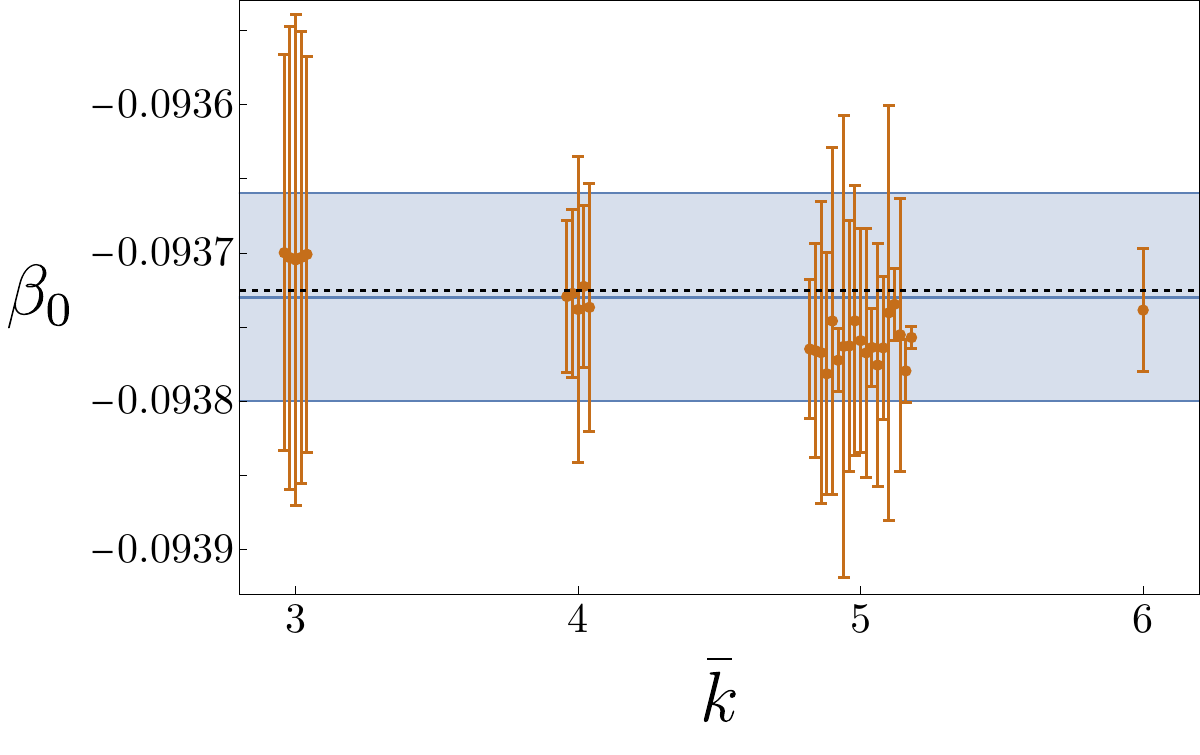}}
  \caption{ \emph{Left panel}: Comparison between the values $\Delta_0(\hat{Q})$ that we have computed for $\hat{Q} \geq 1$ and the leading order behavior in $\hat{Q}$ of the large-charge expansion that is predicted by the computation at NLO of the energy density in flat space (\ref{eq:NLO_flat_result}).  \emph{Right panel}: Estimates of the values for the coefficient $\beta_0$ coming from the fitting procedure. The parameter $\bar{k}$ indicated the value for $k$ in the expression (\ref{eq:fit_form_largeQ}) for which the fit is most stable. The several orange points are the outcomes from the repeated fitting performed on our values for $\Delta_0(\hat{Q})$ with random fluctuation in their error intervals. In order to make the plot more comprehensible and not overlap points and error bars we applied small shifts, the values of $\bar{k}$ are always the nearest integer. The dotted line corresponds to the theoretical value and the blue bar to our final estimate. We refer the reader to appendix \ref{sec:Fitting} for further details.}
  \label{fig:LargeQ}
\end{figure}
of the form $\hat{Q}^{1/2}$ and $\hat{Q}^0$, confirming once again the expectations.
We can now concentrate on the coefficients of this expansion.  
Let us start with $\alpha^{(0)}_{3/2}$.  
Going back to the effective potential $\mathcal{V}^{(0)}_{\text{cyl.}}(\mu)$, this is the  coefficient of the term $ R^2\mu^3$, the only one that survives in the limit  $R\to \infty$. Thus, $\alpha^{(0)}_{3/2}$ is nothing else than the quantity $v^{(0)}$ that appears in the computation done in $\mathbb{R}^3$ (\ref{eq:NLO_flat_result}).\footnote{The factor $\sqrt{4\pi}$ multiplying $v^{(0)}$ in eq.~(\ref{eq:NLO_flat_result}) disappears once the sphere volume factors are taken in account. } In figure \ref{fig:LargeQ}, we plotted the values of $\Delta_0$ computed and the leading behavior coming from the flat space computation.  
Let us now compute the coefficients of the large-$\hat{Q}$ expansion (\ref{eq:largeQ_NLO}) by fitting to our points functions of the type 
\begin{equation} \label{eq:fit_form_largeQ}
h^{[k]}(\hat{Q})=h_{\frac{3}{2}}^{[k]}\hat{Q}^{\frac{3}{2}}+h_{\frac{1}{2}}^{[k]}\hat{Q}^{\frac{1}{2}}+\sum_{i=0}^{k} h^{[k]}_{-\frac{i}{2}} \hat{Q}^{-\frac{i}{2}}\,.
\end{equation}
Once again, our estimation will have an uncertainty that will be the combination of the error on the points, i.e. $\Delta_0$ at fixed $\hat{Q}$, and the truncation on $k$. Too big $k$ will correspond to overfitting,  instead taking $k$ too small will give an error due to the truncation of the expansion. 
We start with $\alpha^{(0)}_{3/2}$, as another check of our numerical computation. 
We find many digits of agreement,
\begin{equation}
 \alpha^{(0)}_{\frac{3}{2}} \equiv v^{(0)}=-0.03530346\dots\,,  \ \qquad \qquad  \alpha^{(0)}_{\frac{3}{2},\text{fit}}=-0.03530348(5)\,.
\end{equation}
\begin{table}[t!]
  \centering
  \begin{tabular}{c||c|c| c | c}
    \hline
    		& $\alpha^{(0)}_{\frac{1}{2}}$	&    $\beta_0$		& $\alpha^{(0)}_{-\frac{1}{2}}$		& $\beta^{(0)}_{-1}$	 \\    \hline
    Theory 	&		& $-0.093725\dots$		& 	& $0.0081\dots$	 \\    \hline
    Fit (this work) 	& $-0.03938216(9)$		& $-0.09373(7)\ \ $		& $0.02364(2)$	& $0.008(2)\ \ $	\\    \hline
  \end{tabular}
  \caption{Coefficient of the large-$\hat{Q}$ expansion of $\Delta_0$. In the first line are reported the values of the coefficients that come from quantum corrections,  i.e. $\beta_0$ and $\beta^{(0)}_{-1}$ from eqs.~(\ref{eq:beta0_Casimir}) and (24) from ref.~\cite{Cuomo:2020rgt}, respectively. 
  In the second one, our results extracted by fitting.  In order to maximize their accuracy,  the estimates for $\alpha^{(0)}_{1/2}$ and $\alpha^{(0)}_{-1/2}$ have been computed last, after inputting in the fit the theoretical expectation for $\beta_0$ and $\beta^{(0)}_{-1}$, given their agreement with their fitted values. Similarly, $\beta^{(0)}_{-1}$ is computed inputting the theoretical expectation for $\beta_0$.}
  \label{tab:largeQ}
\end{table}
From here on, we will input the exact value of $\alpha^{(0)}_{3/2}$ in the ansatz (\ref{eq:fit_form_largeQ}) and focus on the next coefficients in the 
expansion (\ref{eq:largeQ_NLO}).  We find the results reported in the second line of table \ref{tab:largeQ}. First, we concentrated on the coefficients for which we have a theoretical expectation, namely $\beta_0$ and $\beta^{(0)}_{-1}$. The coefficient $\beta_0$ is universal and corresponds to the one-loop Casimir energy of the conformal GB (\ref{eq:beta0_Casimir}).\footnote{Differently from all other coefficients, this term does not have an expansion in $1/N$, thus its labeling without superscript.} The estimation provided to us by the fit is in complete agreement with the theory.  
In figure \ref{fig:LargeQ}, right panel, we report it together with some intermediate findings of the fitting procedure. For more details on the procedure adopted to determine our result and for the numerical estimation of the error, see appendix \ref{sec:Fitting}.  
Then,  following the steps adopted several times,  we constrained in the fit the constant term with the exact value for $\beta_0$ and we looked at $\beta^{(0)}_{-1}$.  
The computation of this coefficient via the fitting procedure is particularly challenging since the term $\hat{Q}^{-1}$ is very subleading in the large-$\hat{Q}$ expansion.  Nevertheless, we find an estimate compatible with the theoretical one \cite{Cuomo:2020rgt} which relates it to the coefficients $c_2,c_3$ at LO, see the last column of table \ref{tab:largeQ}. 
Finally,  inputting the values of all the known coefficients we 
focused on the $\alpha^{(0)}_{1/2}$ and $\alpha^{(0)}_{-1/2}$, for which there is no theoretical expectation.  Our results are reported in the first and third column of table  \ref{tab:largeQ}. 

The values $\alpha^{(0)}_{3/2},\, \alpha^{(0)}_{1/2}$, and $\alpha^{(0)}_{1/2}$ are enough to compute the contribution at NLO at large-$N$ for the Wilson parameters $c_1,$ $c_2$, and $c_4$. We find
\begin{align}
	c_1 &= \frac{N}{12 \pi} +0.00280936\ldots+ O\left(N^{-1}\right)\,,  \nonumber \\
	 c_2 &= \frac{N}{48 \pi} -0.001215796(3)+O\left(N^{-1}\right)\,,\nonumber \\
	 	 c_4 &= -\frac{N}{960 \pi} +0.0005459(4)+O\left(N^{-1}\right)\,.
    \label{eq:Wilson_coeff_NLO}
\end{align}

\begin{figure}[t!]
\centering
  \raisebox{0\height}{\includegraphics[width=0.485\textwidth]{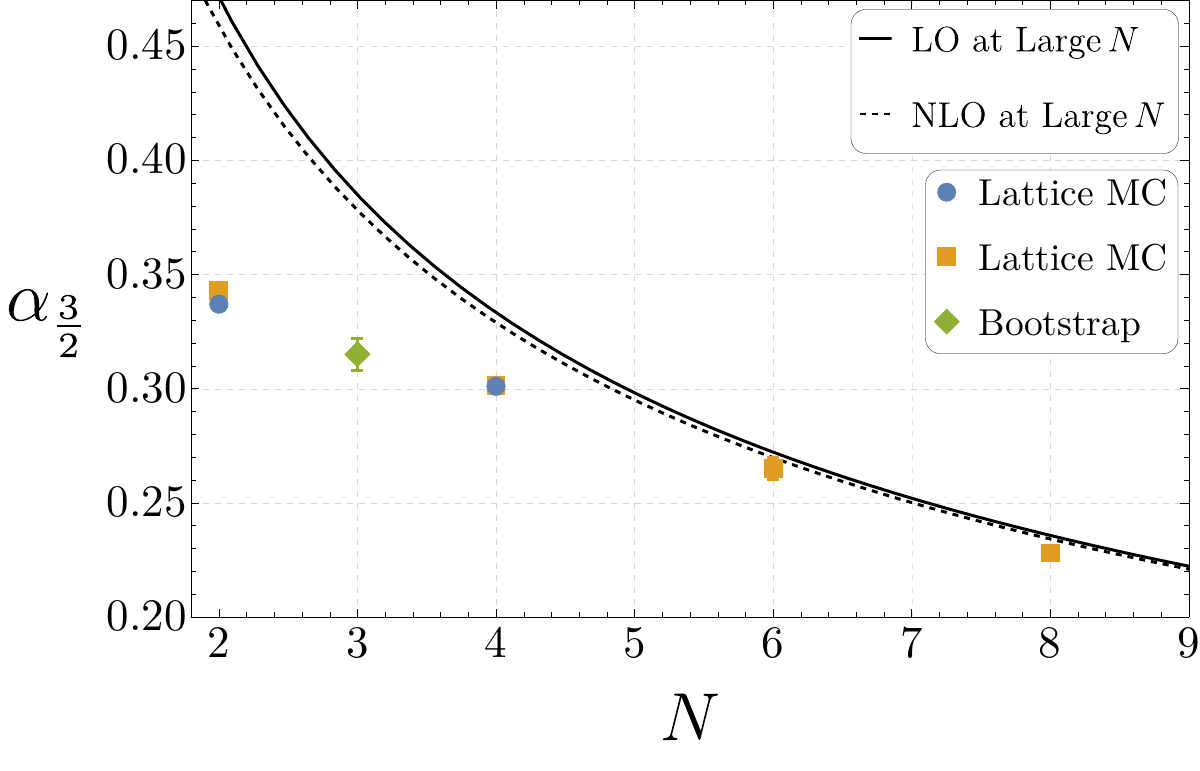}}
  \hspace*{.1cm}
  \raisebox{0\height}{\includegraphics[width=0.485\textwidth]{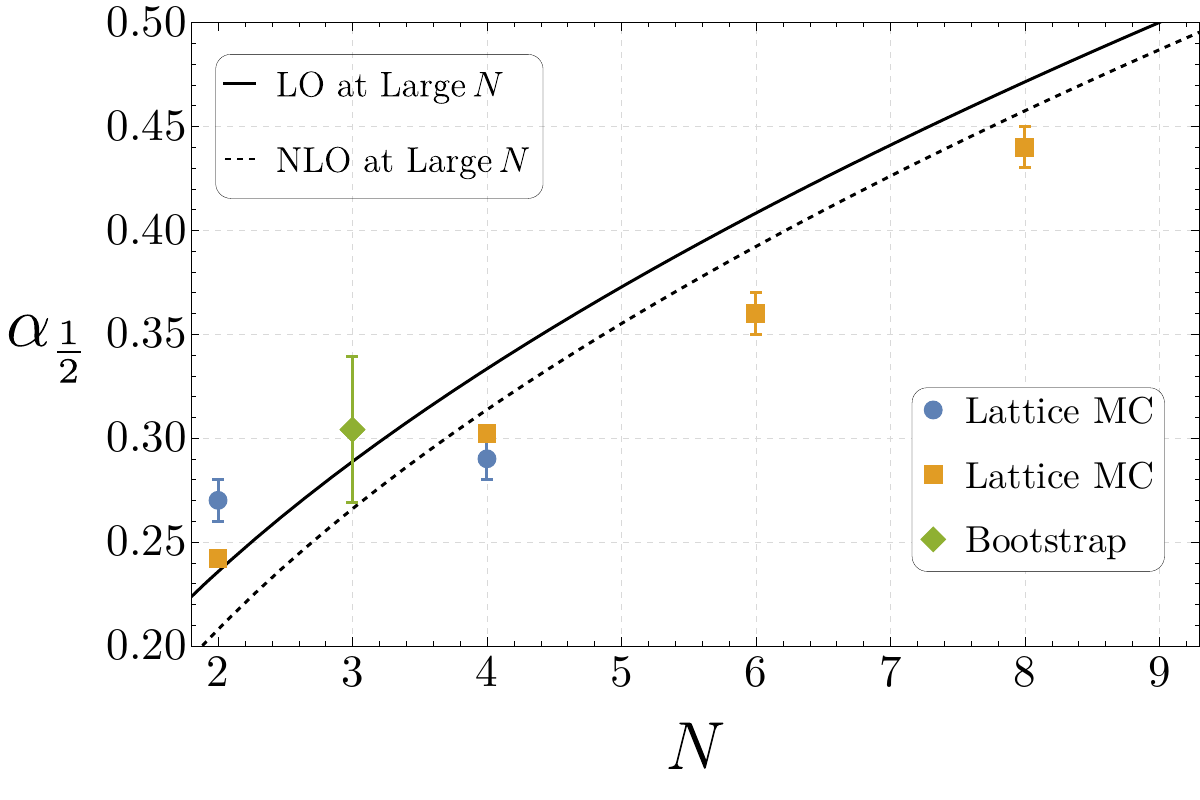}}
  \caption{Comparison between the large-$N$ predictions for $\alpha_{3/2}$ and $\alpha_{1/2}$ at LO and NLO, solid and dashed lines, respectively,  and results at fixed $N$ computed via other techniques in the literature. The blue dots have been obtained via lattice Monte Carlo in refs. \cite{Banerjee:2017fcx,Banerjee:2019jpw}, the orange square via lattice Monte Carlo in ref.~\cite{Singh:2022akp} and the green diamonds via a numerical bootstrap computation in ref.~\cite{Rong:2023owx}.}
  \label{fig:comparison_finiteN}
\end{figure}

To conclude this section, let us compare the coefficients of the large-$\hat{Q}$ expansion at large $N$ at leading and next-to-leading order with values computed at fixed $N$ via other techniques.  In figure \ref{fig:comparison_finiteN}, we plotted the LO and NLO at large $N$ for $\alpha_{3/2}$ and $\alpha_{1/2}$ in the left and right panel, respectively.  
Their expansion is of the form of eq.~(\ref{eq:expansion_N_alpha}). Thus,
\begin{equation}
\alpha_{\frac{3}{2}}=\alpha^{(-1)}_\frac{3}{2} N^{-\frac{1}{2}}+\alpha^{(0)}_\frac{3}{2}N^{-\frac{3}{2}}+\dots\,, \qquad\qquad \alpha_{\frac{1}{2}}=\alpha^{(-1)}_\frac{1}{2} N^{\frac{1}{2}}+\alpha^{(0)}_\frac{1}{2}N^{-\frac{1}{2}}+\dots\,.
\end{equation}
The values at fixed $N$ were obtained in refs.~\cite{Banerjee:2017fcx,Banerjee:2019jpw} via lattice Monte Carlo for the $N=2$ and $N=4$,  in ref.~\cite{Singh:2022akp} via lattice Monte Carlo for the $N=2,4,6,8$ and via a numerical bootstrap computation for $N=3$ in ref.~\cite{Rong:2023owx}. The large-$N$ predictions for the leading coefficient $\alpha_{3/2}$ were already showing a nice agreement improving rapidly with $N$. The NLO correction is very small, but it goes in the right direction. For the coefficient $\alpha_{1/2}$, the NLO correction is notably improving the agreement with the trend of the data, alleviating the mild tension present at the LO as pointed out in ref.~\cite{Singh:2022akp}. To our knowledge, there is no numerical estimate for subleading coefficients like $\alpha_{-1/2}$ to compare to in the present literature.

\section{Conclusion}
\label{sec:conclusion}

In this work we computed the scaling dimension of the lowest-lying large charge operator in the critical $O(N)$ model in the double scaling limit $N,Q \rightarrow \infty,\, \hat{Q} \coloneqq Q/N$ fixed. This was done by adapting the methodology previously used to compute scaling dimensions of monopole operators \cite{Dyer:2015zha, Pufu:2013eda, Pufu:2013vpa, Chester:2021drl, Chester:2017vdh}. Extrapolating our result to the regime $\hat{Q} \ll 1$, we could compare with ordinary large-$N$ perturbation theory, matching the first two known coefficients appearing at $O(1/N, 1/N^2)$ and providing a (numerical) prediction for the $O(1/N^3, 1/N^4)
$. These results are summarized in table \ref{tab:smallQ}. In the regime $\hat{Q} \gg 1$, we obtained a behavior consistent with the large-charge EFT prediction. We were able to extract NLO corrections in the double scaling limit for the first three Wilson coefficients appearing in the EFT, $c_1 , c_2, $ and $c_4$ as well as match the prediction for the coefficients $\beta_0, \, \beta_{-1}$ which are sensible to the corrections to the EFT Goldstone boson dispersion relation. These results are summarized in table \ref{tab:largeQ} and eq.~(\ref{eq:Wilson_coeff_NLO}).

There are a number of relevant directions we leave for future research.
\begin{itemize}
	\item One could apply the same methodology used in this work to compute the NLO correction to scaling dimensions of large-charge operators which correspond to Fermi sphere ground states \cite{Dondi:2022zna}. Since a Fermi liquid EFT construction for this specific problem has not been attempted yet,\footnote{We are aware of work in progress by Jahmall Bersini, Simeon Hellerman, Domenico Orlando, and Susanne Reffert in this direction. It was also put forward the possibility that such Fermi sphere ground states might not be stable at all and it is not clear whether a NLO computation would settle this.} there are no predictions to match, and this computation would lead to a great deal of novel information for these types of operators. 
	\item The extension to operators in the $O(N)$ model in large asymmetric representations or with spin comparable with their $O(N)$ quantum number has not yet been worked out, even at LO in the double scaling limit employed in this work. This is a case in which the usual large-charge EFT does not apply in its simplest form and would provide checks and extra information for the conjectured new EFT phases appearing in these regimes. 
    \item Extending our result to generic $d$ in the range $2<d<4$ will allow matching to results obtained in the $\epsilon$-expansion \cite{Badel:2019oxl}. Notice that the $\epsilon$-result is organized differently from our eq.~\eqref{eq:LargeN_Delta}, and matching our results for $\Delta_0$ would require an extra order compared with what is presently found in the literature. Such a result would be an interesting quantity to compute in its own right. 
    \item A flat-space-only computation for the NLO corrections to the $\hat{\sigma}$-propagators can lead to an estimate of the $\sim N^0$ contribution to the Wilson coefficient $c_3$,  for which the order $ \sim N$ vanishes
     (notice that to compute $c_3$ to this order one needs the NLO determination of the $c_2$ coefficient we estimated numerically \eqref{eq:gamma}).
\end{itemize}

\section*{Acknowledgments}
We thank Jahmall Bersini, Agnese Bissi,  Gabriel Cuomo, Simeon Hellerman, Domenico Orlando, and Susanne Reffert for helpful discussions and comments on the draft. 
We also thank the participants and organizers of the Workshop ``Towards realistic physics at large quantum number" held at Tokyo IPMU in May 2024 where results of this work were presented.  
NAD acknowledges the receipt of the joint grant from the Abdus Salam International Centre for Theoretical Physics (ICTP), Trieste, Italy and INFN, sede centrale Frascati, Italy. 
The work of GS is supported by the Swiss National Science Foundation under grant number 200021~192137.

\appendix

\section{Thermodynamical relations}
\label{sec:termo}
In flat space, where we write $\mathcal{Z} = \exp(- \text{Volume} \times V)$, the charge and energy density of a homogeneous saddle configuration are computed as
\begin{equation}
	\rho = - \frac{\partial}{\partial \mu} V(\mu), \quad \epsilon(\rho) = V(\mu) + \mu \rho \, .
\end{equation}
These formulas are adapted to the cylinder $V \rightarrow V_{\text{cyl.}}$ by restoring $S^{d-1}_R$-volume factors. The same relations hold with $\rho \rightarrow Q,\, \epsilon(\rho) \rightarrow E(Q)$, which are the total charge and energy respectively. 
	
In the large-$N$ expansion we have together with eq.~(\ref{eq:V_general}),
\begin{equation}
	\hat{\rho} \coloneqq \frac{\rho}{N} \qquad \text{and} \qquad \epsilon(\hat{\rho})=N\epsilon^{(-1)}(\hat{\rho})+\epsilon^{(0)}(\hat{\rho})+O\left(N^{-1}\right)
\end{equation}
Let us expand at the next-to-leading order at large-$N$ the Legendre transform
\begin{equation}
\rho = N \underbrace{  \left( -\frac{\partial V^{(-1)}}{\partial \mu} \right) }_{\coloneqq f_0^{-1}(\mu)} + \underbrace{\left(-\frac{\partial V^{(0)}}{\partial \mu} \right)}_{\coloneqq f_1^{-1}(\mu)} + \dots \,.
\end{equation}
The inverse transform is going to be
\begin{equation}
 f_0(\hat{\rho}) = f_0\left(  f_0^{-1}(\mu) + \frac{1}{N} f_1^{-1}(\mu) +\dots \right) = \mu + \frac{1}{N} f_0'\left(f_0^{-1}(\mu)\right) f_1^{-1}(\mu) + \dots
\end{equation}
which results in
\begin{equation}
\mu = f_0(\hat{\rho}) - \frac{1}{N} f_0'(\hat{\rho}) f_1^{-1}\left(f_0(\hat{\rho})\right) + \dots\,.
\end{equation}
Thus, computing the energy density at next-to-leading order one finds
\begin{align}
\epsilon(\hat{\rho}) &= N \left[ V^{(-1)}(f_0(\hat{\rho})) + f_0(\hat{\rho}) \hat{\rho} \right] \nonumber \\
&+ \left[ V^{(0)}(f_0(\hat{\rho})) - \hat{\rho} f_0'(\hat{\rho}) f_1^{-1}\left(f_0(\hat{\rho})\right) -\left. \frac{\partial V^{(-1)}}{\partial \mu} \right|_{f_0(\hat{\rho})}  f_0'(\hat{\rho}) f_1^{-1}(f_0(\hat{\rho}))   \right] + \dots \nonumber\\
&= N \left[ V^{(-1)}(f_0(\hat{\rho})) + f_0(\hat{\rho}) \hat{\rho} \right] +V^{(0)}(f_0(\hat{\rho})) + \dots\,.
\end{align}
This last expression shows that the NLO correction to the energy density is computed by evaluating the NLO effective potential at the chemical potential $\mu=\mu(Q)$ obtained already at LO.

\section{Computational details on $S^1_\beta \times S^{d-1}_R$}
\label{sec:AppSphHarm}

The hyperspherical harmonic \(Y_{\ell m}\) is an eigenfunction of the Laplacian on \(S^{d-1}\) with eigenvalue
\begin{equation}
	-\Delta_{S^{d-1}} Y_{\ell m} (n) = \ell ( \ell + d - 2 ) Y_{\ell m} (n) ,
\end{equation}
where \(\ell = 0, 1, \dots \) and \(m \) is a vector of \( d - 2\) components satisfying $l \geq m_1 \geq m_2 \geq ... \geq m_{d-3} \geq |m_{d-2}|$  where the lowest component \(m_{d-2}\) is associated to the standard \(SO(3)\) quantum number. This is the only component that can have a negative sign. We denote with \(m^*\) the vector with the sign of \(m_{d-2}\) flipped.
This appears in the conjugation property
\begin{equation}
	Y_{\ell m }^* = (-1)^{m_{d-2}} Y_{\ell m^*}  .
\end{equation}
The eigenvalue does not depend on \(m\), and it appears with multiplicity
\begin{equation}\label{eq:laplacian_degenerancy}
	\text{deg}(\ell) = \frac{(d + 2 \ell - 2) \Gamma(d + \ell - 2)}{\Gamma(\ell + 1) \Gamma(d - 1)} \, .
\end{equation}

The harmonics \(Y_{\ell m}\) form an orthonormal basis for \(L^2(S^{d - 1})\), we choose the normalisation such that
\begin{equation}
	\int_{S^{d -1}} Y_{\ell m} Y^{*}_{\ell'  m' } =  \delta_{\ell \ell'} \delta_{m m'}\,, \qquad \Omega_d = \frac{2 \pi^{d/2}}{\Gamma(d/2)}\,.
\end{equation}
We denote as $\langle \ell_1 m_1 | \, \ell_2 m_2 ; \ell_3 m_3 \rangle$ the triple integral of hyperspherical harmonics. In $d=3$ it evaluates to
\begin{align}
	\int_{S^2} Y_{\ell_1 m_1} Y_{\ell_2 m_2} Y_{\ell_3 m_3} &=\sqrt{ \frac{\prod_{i=1}^3(2\ell_i +1)}{4\pi} } \begin{pmatrix}
		\ell_1 & \ell_2 & \ell_3 \\ 0 & 0 & 0 
	\end{pmatrix}
	\begin{pmatrix}
		\ell_1 & \ell_2 & \ell_3 \\ m_1 & m_2 & m_3 
	\end{pmatrix}\,.
\end{align}
In the main text, we also use the following thermal summation formulas
\begin{align}
	\sum_{n_a \in \mathbb{Z}} \frac{1}{\omega_{n_a}^2  + E_{\ell_a}^2}&= \frac{\beta}{2 } \frac{1}{E_{\ell_a}}\,, \nonumber\\ \sum^{n_a + n_b = n}_{n_a,n_b \in \mathbb{Z}} \frac{1}{\omega_{n_a}^2  + E_{\ell_a}^2} \frac{1}{\omega_{n_b}^2  + E_{\ell_b}^2} &= \frac{\beta }{2 } \left[ \frac{1}{E_{\ell_a}} + \frac{1}{E_{\ell_a}} \right] \frac{1}{\omega_n^2 + (E_{\ell_a} + E_{\ell_b} )^2 }\,,
\end{align}
where $\omega_n = 2\pi n / \beta$ are bosonic Matsubara frequencies.

\section{EFT matching}
\label{sec:EFT}

In this section, we relate the $\alpha$-coefficients appearing in the scaling dimension $\Delta(Q)$ (\ref{eq:GS_contribution}) to the Wilson coefficients of the EFT (\ref{eq:EFT}). We extend the computation of ref.~\cite{Cuomo:2020rgt} to one extra order to match the measured coefficient $\alpha_{-1/2}$ to the Wilson coefficient $c_4$. We refer to the same work for the relation between $\beta_{-1}$ and $c_2,c_3$. 
The operators contributing to the powers $Q^{3/2}, Q^{1/2}$ to the large-charge scaling dimension are
\begin{equation}
	S = \int \dd^d x \sqrt{\hat{g}} \left\{  -c_1 + c_2 \hat{\mathcal{R}} - c_3 \hat{\mathcal{R}}_{\mu\nu} \partial_\mu \chi \partial_\nu \chi     \right\}\,,
\end{equation}
where $\hat{g}_{\mu\nu} =  (\partial \chi)^2 g_{\mu\nu} = (-\partial_\rho \chi \partial^\rho \chi) g_{\mu\nu}$. At the next order, one has
\begin{equation}
	S = \int \dd^d x \sqrt{\hat{g}} \left\{  c_4 \hat{\mathcal{R}}^2 + c_5 \hat{\mathcal{W}}^2 + c_6 \hat{\mathcal{E}} + \text{contractions with the $\partial_\mu \chi$}   \right\}
\end{equation}
since $\partial_\nu \chi|_{\text{GS}} = -i \mu \delta_{\nu 0}$ and on $\mathbb{R} \times S^{d-1}$ the Riemann tensor $\mathcal{R}_{\mu\nu\rho\sigma}$ vanishes whenever an index is zero, we can ignore the terms in the contractions if we only look at the ground-state contributions. This holds also for the operator multiplying $c_3$ at the previous order. We can also ignore the Weyl tensor term $\hat{\mathcal{W}}$ since it's conformal invariant and the Gauss-Bonnet term $\hat{\mathcal{E}}$ being a total derivative. We also use
\begin{align}
	\sqrt{\hat{g}} &= (\partial \chi)^d \sqrt{g}\,, \nonumber\\
	\hat{\mathcal{R}} &= (\partial \chi)^{-2} \left\{    \mathcal{R} - 2 (d-1) \Delta \log (\partial \chi) - (d-2)(d-1) \partial_\mu \log (\partial \chi) \partial^\mu \log (\partial \chi) \right\} 
\end{align}
and notice that the in-homogeneous part of the $\mathcal{R}$-Weyl transformation contains double derivatives on $\chi$, which vanish on the ground state. 
Using $\mathcal{R}|_{\mathbb{R} \times S^{d-1}} = (d-2)(d-1) /R^2$ as well as the fact that on the ground state $(\partial \chi)^2 = \mu^2$, one gets in $d=3$
\begin{equation}
	S|_{\text{GS}} = \frac{\beta}{R} 4\pi \Big\{  -c_1 (\mu R)^{3} + 2 c_2  (\mu R) + 4 c_4 (\mu R)^{-1} + \dots  \Big\}\,.
\end{equation}
The ground state action plays the role of the potential in the thermodynamical formulas of appendix \ref{sec:termo}. One computes the total charge as
\begin{equation}
	Q = - \frac{1}{\beta} \frac{\partial}{\partial \mu} S|_{\text{GS}} =4\pi \Big\{ 3 c_1 (\mu R)^{2} - 2 c_2 + 4 c_4 (\mu R)^{-2} + \dots \Big\}\,,
\end{equation}
which is inverted as $\mu = \mu(Q)$ as follows
\begin{equation}
	\mu R = \left( \frac{Q}{12\pi c_1} \right)^{\frac{1}{2}}\left\{  1 + \frac{c_2}{3 c_1} \left( \frac{Q}{12\pi c_1} \right)^{-1} + \frac{12 c_1 c_4 - c_2^2}{18 c_1^2} \left( \frac{Q}{12\pi c_1} \right)^{-2} + \dots  \right\}\,.
\end{equation}
The scaling dimension is computed as $\Delta(Q)|_{\text{GS}}= (\mu R) Q + S|_{\text{GS}}/\beta$ and reads
\begin{align}
	\Delta(Q)|_{\text{GS}} &= \alpha_{\frac{3}{2}} Q^{\frac{3}{2}} + \alpha_{\frac{1}{2}} Q^{\frac{1}{2}} + \alpha_{-\frac{1}{2}} Q^{-\frac{1}{2}} + \dots\,,  \qquad \text{with} \nonumber\\
	\alpha_{\frac{3}{2}} &= \frac{1}{3\sqrt{3 \pi c_1}}, \quad \alpha_{\frac{1}{2}} = \frac{4 c_2 \sqrt{\pi}}{\sqrt{3 c_1}}, \quad \alpha_{-\frac{1}{2}} = \frac{8\sqrt{3 \pi^3}}{3 \sqrt{c_1}} (12 c_1 c_4 + c_2^2) \, .
\end{align}

\section{Asymptotic expansion}
\label{sec:AppAsymp}

Following appendix C of \cite{Dyer:2015zha}, we obtain the following expression for $\Pi_\ell$:
\begin{align} \label{eq:Pi_exp_largelomega}
\Pi_{\ell}\left( \omega, \mu^2-\frac{1}{4}\right) &= \Pi_{\ell}( \omega, 0)-\frac{\Phi^2_{*\text{cyl.}}}{l^2+\omega^2} 
- \frac{\left(\mathcal{V}_{\text{cyl.}}^{(-1)}+\frac{\pi}{4} \Phi^2_{*\text{cyl.}}\right) \omega^2}{\pi\left((l+2)^2+\omega^2\right)\left(l^2+\omega^2\right)\left((l-2)^2+\omega^2\right)} \nonumber \\
&+ \frac{\left(\mathcal{V}_{\text{cyl.}}^{(-1)}+\frac{\pi}{2}\Phi^2_{*\text{cyl.}}\left(8\mu^2-1\right)\right)\left(l+\frac{1}{2}\right)\left(l-\frac{1}{2}\right)}{2\pi\left((l+2)^2+\omega^2\right)\left(l^2+\omega^2\right)\left((l-2)^2+\omega^2\right)}
+ O\left(\left(l^2+\omega^2\right)^{-3}\right)\,,
\end{align}
where 
\begin{equation} \label{eq:Pi0_exp_largelomega}
\Pi_{\ell}( \omega, 0) = \frac{1}{16R}\left(\frac{1}{\sqrt{l^2+\omega^2}} + \frac{\omega^2-l^2}{8\left(l^2+\omega^2\right)^{\frac{5}{2}}} + \frac{11\omega^4-62\omega^2 l^2+11l^4}{128\left(l^2+\omega^2\right)^{\frac{9}{2}}} +O\left(\left(l^2+\omega^2\right)^{-\frac{7}{2}}\right)\right)\,.
\end{equation}
In the expressions above, we introduced a short-hand notation $l\coloneqq \left(\ell+\frac{1}{2}\right)/R$. In the eqs.~(\ref{eq:Pi_exp_largelomega}) and (\ref{eq:Pi0_exp_largelomega}), we set $R=1$ to lighten the notation.

\section{Fitting details}
\label{sec:Fitting}

In this appendix, we provide details on the procedure that we have adopted to estimate the coefficients of the small- and large-$\hat{Q}$ expansions 
starting from the numerical computation of $\Delta_0(\hat{Q})$.
First of all, we choose a set of values for $\hat{Q}$ for which to compute $\Delta_0$. We have to take into consideration that, as mentioned at the end of section \ref{subsec:Numercal}, it is a considerably more costly computation as we take bigger $\hat{Q}$. Then, we compute $\Delta_0$ for these values. Our points will clearly have an uncertainty, 
which is derived from the approximations presented in section \ref{subsec:Numercal}.
Next, we fit the truncated theoretical expansion to these values.  For example in the case of  large-$\hat{Q}$,  we start from  
\begin{equation}
h^{[k]}(\hat{Q})=h_{\frac{3}{2}}^{[k]}\hat{Q}^{\frac{3}{2}}+h_{\frac{1}{2}}^{[k]}\hat{Q}^{\frac{1}{2}}+\sum_{i=0}^{k} h^{[k]}_{-\frac{i}{2}} \hat{Q}^{-\frac{i}{2}}\,
\end{equation}
and find numerical values of the parameters $h_{i}^{[k]}$ that make the function $h^{[k]}(\hat{Q})$ give a best fit to our points as a function of $\hat{Q}$. For the moment, let us neglect the uncertainty, and we will specify shortly how we accounted for it.  Let us focus on a specific coefficient $h_{j}^{[k]}$ that we are interested in. There is very often an interval of stability for $k$ for which the difference between the $h_{j}^{[k]}$ are considerably smaller than the rest. We take as estimate of the result and its error as follows
\begin{equation}\label{eq:fit_estimate}
h^{\text{fit}}_{j} =\frac{1}{2} \left(h_{j}^{[\bar{k}]}+h_{j}^{[\bar{k}+1]}\right)\,, \qquad
\delta h^{\text{fit}}_{j} = 4\left(h_{j}^{[\bar{k}]}-h_{j}^{[\bar{k}+1]}\right)\,,
\end{equation}
where $\bar{k}$ is the value for which the difference $h_{j}^{[\bar{k}]}-h_{j}^{[\bar{k}+1]}$ is the minimum one.  The factor 4 is to account for possible underestimation of the error since we noticed that often all the values $h_{j}^{[k]}$ for $k$ in the region of stability get to be compatible with $h^{\text{fit}}_{j} \pm \delta h^{\text{fit}}_{j}$, as defined in eq.~(\ref{eq:fit_estimate}).
To account for the uncertainty of the points, i.e. the values of $\Delta_0(\hat{Q})$ we computed, we calculated 30 times $h^{\text{fit}}_{j} \pm \delta h^{\text{fit}}_{j}$ from points with fluctuations randomly generated with a flat distribution of range equal to their uncertainty.  We checked if most of the results were compatible. 
We computed the mean and propagated the errors 
to obtain the final result for the given set of points with uncertainty. 
Moreover, as a check of the stability of the result, we checked whether this estimate is stable when we remove one or more points from our set and averaged between the results.
The repeated execution of the fitting procedure tests the goodness of the algorithms used for estimating $h^{\text{fit}}_{j} \pm \delta h^{\text{fit}}_{j}$. It also enables the calculation of arithmetic means, which helps compensate for over- or under-estimated uncertainties in the estimates.

\bibliographystyle{JHEP}
\bibliography{refs}

\providecommand{\href}[2]{#2}\begingroup\raggedright\begin{thebibliography}{10}

\bibitem{Fitzpatrick:2012yx}
A.L.~Fitzpatrick, J.~Kaplan, D.~Poland and D.~Simmons-Duffin, \emph{{The
  Analytic Bootstrap and AdS Superhorizon Locality}},
  \href{https://doi.org/10.1007/JHEP12(2013)004}{\emph{JHEP} {\bfseries 12}
  (2013) 004} [\href{https://arxiv.org/abs/1212.3616}{{\ttfamily 1212.3616}}].

\bibitem{Komargodski:2012ek}
Z.~Komargodski and A.~Zhiboedov, \emph{{Convexity and Liberation at Large
  Spin}}, \href{https://doi.org/10.1007/JHEP11(2013)140}{\emph{JHEP} {\bfseries
  11} (2013) 140} [\href{https://arxiv.org/abs/1212.4103}{{\ttfamily
  1212.4103}}].

\bibitem{Alday:2015eya}
L.F.~Alday, A.~Bissi and T.~Lukowski, \emph{{Large spin systematics in CFT}},
  \href{https://doi.org/10.1007/JHEP11(2015)101}{\emph{JHEP} {\bfseries 11}
  (2015) 101} [\href{https://arxiv.org/abs/1502.07707}{{\ttfamily
  1502.07707}}].

\bibitem{Hellerman:2015nra}
S.~Hellerman, D.~Orlando, S.~Reffert and M.~Watanabe, \emph{{On the CFT
  Operator Spectrum at Large Global Charge}},
  \href{https://doi.org/10.1007/JHEP12(2015)071}{\emph{JHEP} {\bfseries 12}
  (2015) 071} [\href{https://arxiv.org/abs/1505.01537}{{\ttfamily
  1505.01537}}].

\bibitem{Monin:2016jmo}
A.~Monin, D.~Pirtskhalava, R.~Rattazzi and F.K.~Seibold, \emph{{Semiclassics,
  Goldstone Bosons and CFT data}},
  \href{https://doi.org/10.1007/JHEP06(2017)011}{\emph{JHEP} {\bfseries 06}
  (2017) 011} [\href{https://arxiv.org/abs/1611.02912}{{\ttfamily
  1611.02912}}].

\bibitem{Pappadopulo:2012jk}
D.~Pappadopulo, S.~Rychkov, J.~Espin and R.~Rattazzi, \emph{{OPE Convergence in
  Conformal Field Theory}},
  \href{https://doi.org/10.1103/PhysRevD.86.105043}{\emph{Phys. Rev. D}
  {\bfseries 86} (2012) 105043}
  [\href{https://arxiv.org/abs/1208.6449}{{\ttfamily 1208.6449}}].

\bibitem{Mukhametzhanov:2018zja}
B.~Mukhametzhanov and A.~Zhiboedov, \emph{{Analytic Euclidean Bootstrap}},
  \href{https://doi.org/10.1007/JHEP10(2019)270}{\emph{JHEP} {\bfseries 10}
  (2019) 270} [\href{https://arxiv.org/abs/1808.03212}{{\ttfamily
  1808.03212}}].

\bibitem{Benjamin:2023qsc}
N.~Benjamin, J.~Lee, H.~Ooguri and D.~Simmons-Duffin, \emph{{Universal
  asymptotics for high energy CFT data}},
  \href{https://doi.org/10.1007/JHEP03(2024)115}{\emph{JHEP} {\bfseries 03}
  (2024) 115} [\href{https://arxiv.org/abs/2306.08031}{{\ttfamily
  2306.08031}}].

\bibitem{Benjamin:2024kdg}
N.~Benjamin, J.~Lee, S.~Pal, D.~Simmons-Duffin and Y.~Xu, \emph{{Angular
  fractals in thermal QFT}},
  \href{https://arxiv.org/abs/2405.17562}{{\ttfamily 2405.17562}}.

\bibitem{Cuomo:2020rgt}
G.~Cuomo, \emph{{A note on the large charge expansion in 4d CFT}},
  \href{https://doi.org/10.1016/j.physletb.2020.136014}{\emph{Phys. Lett. B}
  {\bfseries 812} (2021) 136014}
  [\href{https://arxiv.org/abs/2010.00407}{{\ttfamily 2010.00407}}].

\bibitem{Dondi:2022wli}
N.~Dondi, I.~Kalogerakis, R.~Moser, D.~Orlando and S.~Reffert, \emph{{Spinning
  correlators in large-charge CFTs}},
  \href{https://doi.org/10.1016/j.nuclphysb.2022.115928}{\emph{Nucl. Phys. B}
  {\bfseries 983} (2022) 115928}
  [\href{https://arxiv.org/abs/2203.12624}{{\ttfamily 2203.12624}}].

\bibitem{Monin:2016bwf}
A.~Monin, \emph{{Partition function on spheres: How to use zeta function
  regularization}},
  \href{https://doi.org/10.1103/PhysRevD.94.085013}{\emph{Phys. Rev. D}
  {\bfseries 94} (2016) 085013}
  [\href{https://arxiv.org/abs/1607.06493}{{\ttfamily 1607.06493}}].

\bibitem{Hellerman:2017efx}
S.~Hellerman, N.~Kobayashi, S.~Maeda and M.~Watanabe, \emph{{A Note on
  Inhomogeneous Ground States at Large Global Charge}},
  \href{https://doi.org/10.1007/JHEP10(2019)038}{\emph{JHEP} {\bfseries 10}
  (2019) 038} [\href{https://arxiv.org/abs/1705.05825}{{\ttfamily
  1705.05825}}].

\bibitem{Hellerman:2018sjf}
S.~Hellerman, N.~Kobayashi, S.~Maeda and M.~Watanabe, \emph{{Observables in
  inhomogeneous ground states at large global charge}},
  \href{https://doi.org/10.1007/JHEP08(2021)079}{\emph{JHEP} {\bfseries 08}
  (2021) 079} [\href{https://arxiv.org/abs/1804.06495}{{\ttfamily
  1804.06495}}].

\bibitem{Cuomo:2019ejv}
G.~Cuomo, \emph{{Superfluids, vortices and spinning charged operators in 4d
  CFT}}, \href{https://doi.org/10.1007/JHEP02(2020)119}{\emph{JHEP} {\bfseries
  02} (2020) 119} [\href{https://arxiv.org/abs/1906.07283}{{\ttfamily
  1906.07283}}].

\bibitem{Cuomo:2022kio}
G.~Cuomo and Z.~Komargodski, \emph{{Giant Vortices and the Regge Limit}},
  \href{https://doi.org/10.1007/JHEP01(2023)006}{\emph{JHEP} {\bfseries 01}
  (2023) 006} [\href{https://arxiv.org/abs/2210.15694}{{\ttfamily
  2210.15694}}].

\bibitem{Gaume:2020bmp}
L.A.~Gaum\'e, D.~Orlando and S.~Reffert, \emph{{Selected topics in the large
  quantum number expansion}},
  \href{https://doi.org/10.1016/j.physrep.2021.08.001}{\emph{Phys. Rept.}
  {\bfseries 933} (2021) 1} [\href{https://arxiv.org/abs/2008.03308}{{\ttfamily
  2008.03308}}].

\bibitem{Komargodski:2021zzy}
Z.~Komargodski, M.~Mezei, S.~Pal and A.~Raviv-Moshe, \emph{{Spontaneously
  broken boosts in CFTs}},
  \href{https://doi.org/10.1007/JHEP09(2021)064}{\emph{JHEP} {\bfseries 09}
  (2021) 064} [\href{https://arxiv.org/abs/2102.12583}{{\ttfamily
  2102.12583}}].

\bibitem{Dondi:2022zna}
N.~Dondi, S.~Hellerman, I.~Kalogerakis, R.~Moser, D.~Orlando and S.~Reffert,
  \emph{{Fermionic CFTs at large charge and large N}},
  \href{https://doi.org/10.1007/JHEP08(2023)180}{\emph{JHEP} {\bfseries 08}
  (2023) 180} [\href{https://arxiv.org/abs/2211.15318}{{\ttfamily
  2211.15318}}].

\bibitem{Badel:2019oxl}
G.~Badel, G.~Cuomo, A.~Monin and R.~Rattazzi, \emph{{The Epsilon Expansion
  Meets Semiclassics}},
  \href{https://doi.org/10.1007/JHEP11(2019)110}{\emph{JHEP} {\bfseries 11}
  (2019) 110} [\href{https://arxiv.org/abs/1909.01269}{{\ttfamily
  1909.01269}}].

\bibitem{Antipin:2020abu}
O.~Antipin, J.~Bersini, F.~Sannino, Z.-W.~Wang and C.~Zhang, \emph{{Charging
  the $O(N)$ model}},
  \href{https://doi.org/10.1103/PhysRevD.102.045011}{\emph{Phys. Rev. D}
  {\bfseries 102} (2020) 045011}
  [\href{https://arxiv.org/abs/2003.13121}{{\ttfamily 2003.13121}}].

\bibitem{Antipin:2020rdw}
O.~Antipin, J.~Bersini, F.~Sannino, Z.-W.~Wang and C.~Zhang, \emph{{Charging
  non-Abelian Higgs theories}},
  \href{https://doi.org/10.1103/PhysRevD.102.125033}{\emph{Phys. Rev. D}
  {\bfseries 102} (2020) 125033}
  [\href{https://arxiv.org/abs/2006.10078}{{\ttfamily 2006.10078}}].

\bibitem{Antipin:2021jiw}
O.~Antipin, J.~Bersini, F.~Sannino, Z.-W.~Wang and C.~Zhang, \emph{{More on the
  cubic versus quartic interaction equivalence in the $O(N)$ model}},
  \href{https://doi.org/10.1103/PhysRevD.104.085002}{\emph{Phys. Rev. D}
  {\bfseries 104} (2021) 085002}
  [\href{https://arxiv.org/abs/2107.02528}{{\ttfamily 2107.02528}}].

\bibitem{Antipin:2022naw}
O.~Antipin, J.~Bersini and P.~Panopoulos, \emph{{Yukawa interactions at large
  charge}}, \href{https://doi.org/10.1007/JHEP10(2022)183}{\emph{JHEP}
  {\bfseries 10} (2022) 183}
  [\href{https://arxiv.org/abs/2208.05839}{{\ttfamily 2208.05839}}].

\bibitem{Alvarez-Gaume:2019biu}
L.~Alvarez-Gaume, D.~Orlando and S.~Reffert, \emph{{Large charge at large N}},
  \href{https://doi.org/10.1007/JHEP12(2019)142}{\emph{JHEP} {\bfseries 12}
  (2019) 142} [\href{https://arxiv.org/abs/1909.02571}{{\ttfamily
  1909.02571}}].

\bibitem{Giombi:2020enj}
S.~Giombi and J.~Hyman, \emph{{On the large charge sector in the critical O(N)
  model at large N}},
  \href{https://doi.org/10.1007/JHEP09(2021)184}{\emph{JHEP} {\bfseries 09}
  (2021) 184} [\href{https://arxiv.org/abs/2011.11622}{{\ttfamily
  2011.11622}}].

\bibitem{Dondi:2021buw}
N.~Dondi, I.~Kalogerakis, D.~Orlando and S.~Reffert, \emph{{Resurgence of the
  large-charge expansion}},
  \href{https://doi.org/10.1007/JHEP05(2021)035}{\emph{JHEP} {\bfseries 05}
  (2021) 035} [\href{https://arxiv.org/abs/2102.12488}{{\ttfamily
  2102.12488}}].

\bibitem{Cuomo:2023mxg}
G.~Cuomo, J.M.V.P.~Lopes, J.~Matos, J.~Oliveira and J.~Penedones,
  \emph{{Numerical tests of the large charge expansion}},
  \href{https://doi.org/10.1007/JHEP05(2024)161}{\emph{JHEP} {\bfseries 05}
  (2024) 161} [\href{https://arxiv.org/abs/2305.00499}{{\ttfamily
  2305.00499}}].

\bibitem{Banerjee:2017fcx}
D.~Banerjee, S.~Chandrasekharan and D.~Orlando, \emph{{Conformal dimensions via
  large charge expansion}},
  \href{https://doi.org/10.1103/PhysRevLett.120.061603}{\emph{Phys. Rev. Lett.}
  {\bfseries 120} (2018) 061603}
  [\href{https://arxiv.org/abs/1707.00711}{{\ttfamily 1707.00711}}].

\bibitem{Banerjee:2019jpw}
D.~Banerjee, S.~Chandrasekharan, D.~Orlando and S.~Reffert, \emph{{Conformal
  dimensions in the large charge sectors at the O(4) Wilson-Fisher fixed
  point}}, \href{https://doi.org/10.1103/PhysRevLett.123.051603}{\emph{Phys.
  Rev. Lett.} {\bfseries 123} (2019) 051603}
  [\href{https://arxiv.org/abs/1902.09542}{{\ttfamily 1902.09542}}].

\bibitem{Banerjee:2021bbw}
D.~Banerjee and S.~Chandrasekharan, \emph{{Subleading conformal dimensions at
  the O(4) Wilson-Fisher fixed point}},
  \href{https://doi.org/10.1103/PhysRevD.105.L031507}{\emph{Phys. Rev. D}
  {\bfseries 105} (2022) L031507}
  [\href{https://arxiv.org/abs/2111.01202}{{\ttfamily 2111.01202}}].

\bibitem{Rong:2023owx}
J.~Rong and N.~Su, \emph{{From O(3) to Cubic CFT: Conformal Perturbation and
  the Large Charge Sector}},
  \href{https://arxiv.org/abs/2311.00933}{{\ttfamily 2311.00933}}.

\bibitem{Dyer:2015zha}
E.~Dyer, M.~Mezei, S.S.~Pufu and S.~Sachdev, \emph{{Scaling dimensions of
  monopole operators in the $ \mathbb{C}{\mathrm{\mathbb{P}}}^{N_b-1} $ theory
  in 2 $+$ 1 dimensions}},
  \href{https://doi.org/10.1007/JHEP03(2016)111}{\emph{JHEP} {\bfseries 06}
  (2015) 037} [\href{https://arxiv.org/abs/1504.00368}{{\ttfamily
  1504.00368}}].

\bibitem{Pufu:2013eda}
S.S.~Pufu and S.~Sachdev, \emph{{Monopoles in 2 + 1-dimensional conformal field
  theories with global U(1) symmetry}},
  \href{https://doi.org/10.1007/JHEP09(2013)127}{\emph{JHEP} {\bfseries 09}
  (2013) 127} [\href{https://arxiv.org/abs/1303.3006}{{\ttfamily 1303.3006}}].

\bibitem{Pufu:2013vpa}
S.S.~Pufu, \emph{{Anomalous dimensions of monopole operators in
  three-dimensional quantum electrodynamics}},
  \href{https://doi.org/10.1103/PhysRevD.89.065016}{\emph{Phys. Rev. D}
  {\bfseries 89} (2014) 065016}
  [\href{https://arxiv.org/abs/1303.6125}{{\ttfamily 1303.6125}}].

\bibitem{Chester:2021drl}
S.M.~Chester, \emph{{Anomalous dimensions of monopole operators in scalar
  QED$_{3}$ with Chern-Simons term}},
  \href{https://doi.org/10.1007/JHEP07(2021)034}{\emph{JHEP} {\bfseries 07}
  (2021) 034} [\href{https://arxiv.org/abs/2102.07377}{{\ttfamily
  2102.07377}}].

\bibitem{Chester:2017vdh}
S.M.~Chester, L.V.~Iliesiu, M.~Mezei and S.S.~Pufu, \emph{{Monopole Operators
  in $U(1)$ Chern-Simons-Matter Theories}},
  \href{https://doi.org/10.1007/JHEP05(2018)157}{\emph{JHEP} {\bfseries 05}
  (2018) 157} [\href{https://arxiv.org/abs/1710.00654}{{\ttfamily
  1710.00654}}].

\bibitem{DeLaFuente:2018uee}
A.~De~La~Fuente, \emph{{The large charge expansion at large $N$}},
  \href{https://doi.org/10.1007/JHEP08(2018)041}{\emph{JHEP} {\bfseries 08}
  (2018) 041} [\href{https://arxiv.org/abs/1805.00501}{{\ttfamily
  1805.00501}}].

\bibitem{Marino:2021six}
M.~Marino, R.~Miravitllas and T.~Reis, \emph{{Testing the Bethe ansatz with
  large N renormalons}},
  \href{https://doi.org/10.1140/epjs/s11734-021-00252-4}{\emph{Eur. Phys. J.
  ST} {\bfseries 230} (2021) 2641}
  [\href{https://arxiv.org/abs/2102.03078}{{\ttfamily 2102.03078}}].

\bibitem{Badel:2019khk}
G.~Badel, G.~Cuomo, A.~Monin and R.~Rattazzi, \emph{{Feynman diagrams and the
  large charge expansion in $3-\varepsilon$ dimensions}},
  \href{https://doi.org/10.1016/j.physletb.2020.135202}{\emph{Phys. Lett. B}
  {\bfseries 802} (2020) 135202}
  [\href{https://arxiv.org/abs/1911.08505}{{\ttfamily 1911.08505}}].

\bibitem{Lang:1992zw}
K.~Lang and W.~Ruhl, \emph{{The Critical O(N) sigma model at dimensions 2
  \ensuremath{<} d \ensuremath{<} 4: Fusion coefficients and anomalous
  dimensions}}, \href{https://doi.org/10.1016/0550-3213(93)90417-N}{\emph{Nucl.
  Phys. B} {\bfseries 400} (1993) 597}.

\bibitem{Derkachov:1997ch}
S.E.~Derkachov and A.N.~Manashov, \emph{{The Simple scheme for the calculation
  of the anomalous dimensions of composite operators in the 1/N expansion}},
  \href{https://doi.org/10.1016/S0550-3213(98)00103-5}{\emph{Nucl. Phys. B}
  {\bfseries 522} (1998) 301}
  [\href{https://arxiv.org/abs/hep-th/9710015}{{\ttfamily hep-th/9710015}}].

\bibitem{Singh:2022akp}
H.~Singh, \emph{{Large-charge conformal dimensions at the $O(N)$ Wilson-Fisher
  fixed point}},  \href{https://arxiv.org/abs/2203.00059}{{\ttfamily
  2203.00059}}.

\end{thebibliography}\endgroup

\end{document}